\documentclass{aastex}
\usepackage{natbib,epsfig,emulateapj5,rotate,color}
\tightenlines

\begin{document}
\newcommand{\be}{\begin{eqnarray}}
\newcommand{\ee}{\end{eqnarray}}
\newcommand{\hii}{H{\sc ii }}
\newcommand{\hi}{H{\sc i }}
\newcommand{\etal}{et al.}
\def\llbdm{{\cal L}_{\rm \ell,b,\DM}}
\def\lpsrd{{\cal L}_{\rm psr,d}}
\def\lpsrs{{\cal L}_{\rm psr,s}}
\def\lgals{{\cal L}_{\rm gal,s}}
\def\lxgals{{\cal L}_{\rm xgal,s}}
\def\like{{\cal L}}
\newcommand{\lampsrd}{{\Lambda}_{\rm psr,d}}
\newcommand{\lampsrs}{{\Lambda}_{\rm psr,s}}
\newcommand{\lamgals}{{\Lambda}_{\rm gal,s}}
\newcommand{\lamxgals}{{\Lambda}_{\rm xgal,s}}
\def\lam{{\Lambda}}
\def\dl{{\rm D_L}}
\def\du{{\rm D_U}}
\def\norm{{\cal N}}
\def\npsrd{{N_{\rm psr,d}}}
\def\dhat{{\hat D}}
\def\DM{{\rm DM}}
\def\RM{{\rm RM}}
\def\SM{{\rm SM}}
\def\EM{{\rm EM}}
\def\DMinfty{{\rm DM}_{\infty}}

\def\SMxgal{{\rm SM}_{\theta,x}}
\def\SMgal{{\rm SM}_{\theta,g}}
\def\SMtau{{\rm SM}_{\tau}}

\def\smun{{kpc \,\, m^{-20/3}}} 
\def\cnsq{{C_n^2}} 
\def\nbar{{\overline{n}_e}}
\def\dne{{\delta n_e}}
\def\pne{{P_{\delta n_e}}}

\def\dnud{{\Delta\nu_{\rm d}}}		
\def\dtd{{\Delta t_{\rm d}}}		
\def\taud{{\tau_d}}			
\def\csm{{C_{\rm SM}}}
\def\narms{{N_{\rm arms}}}
\def\nclumps{{N_{\rm clumps}}}

\def\nparms{{N_{\rm parms}}}
\def\nmc{{N_{\rm MC}}}

\def\xvec{{\bf x}}

\def\wiss{{W_{\rm ISS}}}
\def\wc{{W_{\rm C}}}
\def\wdiss{{W_{\rm D,ISS}}}
\def\wdpm{{W_{\rm D,PM}}}

\newcommand{\nec}{{n_e}_c}
\newcommand{\Fc}{{F}_c}
\newcommand{\dc}{{d}_c}

\newcommand{\sgra}{\mbox{Sgr~A${}^*$}}
\newcommand{\dgc}{\ensuremath{D_{\mathrm{GC}}}}
\newcommand{\delgc}{\ensuremath{\Delta_{\mathrm{GC}}}}
\newcommand{\rsun}{\ensuremath{\mathrm{R}_\odot}}
\newcommand{\wlism}{\ensuremath{w_{\rm LISM}}}
\newcommand{\wlhb}{\ensuremath{W_{\rm LHB}}}
\newcommand{\wlsb}{\ensuremath{W_{\rm LSB}}}
\newcommand{\wldr}{\ensuremath{W_{\rm LDR}}}
\newcommand{\halpha}{\ensuremath{\rm H\alpha}}
\newcommand{\nelism}{n_{\rm LISM}}
\newcommand{\neldr}{n_{\rm LDR}}
\newcommand{\nelsb}{n_{\rm LSB}}
\newcommand{\nelhb}{n_{\rm LHB}}
\newcommand{\Flhb}{F_{\rm LHB}}
\newcommand{\Fldr}{F_{\rm LDR}}
\newcommand{\Flsb}{F_{\rm LSB}}
\newcommand{\Flism}{F_{\rm LISM}}
\newcommand{\negal}{n_{\rm GAL}}
\newcommand{\negc}{n_{\rm GC}}
\newcommand{\ok}{$\surd$}

\newcommand{\blc}{B_{\rm LC}}
\newcommand{\IPprime}{IP$^{\prime}$}

\newcommand{\Tsys}{T_{\rm sys}}
\newcommand{\Tsyso}{T_{\rm sys_0}}
\newcommand{\TCN}{T_{\rm CN}}
\newcommand{\Ssys}{S_{\rm sys}}
\newcommand{\Ssyso}{S_{\rm sys_0}}
\newcommand{\SCN}{S_{\rm CN}}
\newcommand{\DCN}{D_{\rm CN}}
\newcommand{\DNeb}{D_{\rm Neb}}
\def\veffvec{{\bf V_{\rm eff}}}
\def\veffperpvec{{\bf V_{\rm eff,\perp}}}
\def\veff{{\rm V_{\rm eff}}}

\def\kms{{km s$^{-1}$}}

\def\vobsperp{{V_{\rm obs, \perp}}}
\def\vismperp{{ {V_{\rm m}}_\perp}}

\def\vp{{V_{\rm p}}}
\def\vobs{{V_{\rm obs}}}
\def\vism{{V_{\rm m}}}

\def\vpvec{{\bf V_{\rm p}}}
\def\vobsvec{{\bf V_{\rm obs}}}
\def\vismvec{{\bf V_{\rm m}}}

\def\vpperpvec{{\bf   {V_{\rm p}}_\perp}}
\def\vobsperpvec{{\bf {V_{\rm obs}}_\perp}}
\def\vismperpvec{{\bf {V_{\rm m}}_\perp}}

\def\viss{{V_{\rm ISS}}}
\def\visssq{{V^2_{\rm ISS}}}
\def\vissu{{V_{\rm ISS, u}}}
\def\vissIndu{{V_{\rm ISS, \alpha, u}}}
\def\vissKolu{{V_{\rm ISS, 5/3, u}}}    
\def\vissSQu{{V_{\rm ISS, 2, u}}}       
\def\aiss{{A_{\rm ISS}}}
\def\aissu{{A_{\rm ISS,u}}}
\def\aissIndu{{A_{\rm ISS,\alpha,u}}}
\def\aissKolu{{A_{\rm ISS, 5/3, u}}}    
\def\DissKolu{{D_{\rm ISS, 5/3, u}}}    
\def\ds{{D_{\rm s}}}
\def\dds{{D - D_{\rm s}}}
\def\ci{{C_1}}
\def\c1u{{C_{1,u}}}
\def\c1Kolu{{C_{1,5/3,u}}}

\def\Deltavpsperpvec{{\Delta {\bf V_{\rm pf}}_{\perp}   }}
\def\Deltavpsperp{{\Delta {V_{\rm pf}}_{\perp}   }}
\def\vfilperpvec{{ {\bf V_{\rm fil}}_{\perp}   }}

\title{The Brightest Pulses in the Universe: \\
Multifrequency Observations of the Crab Pulsar's Giant Pulses}

\smallskip
\author{J. M. Cordes}
\affil{Astronomy Department and NAIC, Cornell University,
 Ithaca, NY~~14853\\ cordes@astro.cornell.edu}
\medskip
\author{N. D. R. Bhat}
\affil{Massachusetts Institute of Technology, Haystack Observatory, Westford, MA 01886 \\
rbhat@haystack.mit.edu}
\medskip
\author{T. H. Hankins}
\affil{Physics Department, New Mexico Institute of Mining \& Technology \\
thankins@nrao.edu}
\medskip
\author{M. A. McLaughlin}
\affil{Jodrell Bank Observatory, University of Manchester, Macclesfield, 
   Cheshire, SK11 9DL, UK \\
   mclaughl@jb.man.ac.uk}
\medskip
\author{J. Kern}
\affil{Physics Department, New Mexico Institute of Mining \& Technology \\
jkern@nrao.edu}
\bigskip
\begin{abstract}
We analyze the Crab pulsar at ten frequencies from 0.43 to 8.8 GHz 
using data obtained at the
Arecibo Observatory  and report the spectral dependence of 
all pulse components and the rate of occurrence of 
large-amplitude `giant' pulses.
Giant pulses 
occur only in the main-and-interpulse components that are manifest
from radio frequencies to gamma-ray energies 
(known as  the `P1' and `P2' components in the high-energy literature).    
Individual giant pulses reach
brightness temperatures of at least $10^{32}$K in our data, which
do not resolve the narrowest pulses, and are known to reach
$10^{37}$K in nanosecond-resolution observations (Hankins \etal\ 2003).
The Crab pulsar's pulses are therefore the brightest known
in the observable universe.  As such, they represent an important
milestone for theories of the pulsar emission mechanism to explain.
In addition, their short durations allow them to serve
as especially sensitive probes of the Crab Nebula and the interstellar
medium.
We identify and quantify frequency structure in individual
giant pulses using a scintillating, amplitude-modulated, polarized
shot-noise model (SAMPSN). 
The frequency structure associated with multipath propagation
decorrelates on a time scale $\sim 25$  sec at 1.5 GHz.
To produce this time scale requires multipath propagation
to be strongly influenced by material within the Crab Nebula.
We also show that some frequency structure decorrelates rapidly,
on time scales less than one spin period, as would be expected
from the shot-noise pattern of nanosecond duration pulses emitted
by the pulsar.
We discuss the detectability of
individual giant pulses as a function of frequency and provenance.
Taking into account the Crab pulsar's locality inside a bright supernova
remnant, we conclude that the brightest pulse in a typical
1-hour observation would be most easily detectable in our lowest
frequency band (0.43 GHz) to a distance $\sim 1.6$ Mpc
at 5$\sigma$.  We also discuss the detection of such pulses
using future instruments such as LOFAR and the SKA.
\end{abstract}
\keywords{Crab pulsar, Crab Nebula, giant pulses, interstellar medium, 
intergalactic medium}


\section{Introduction}\label{sec:intro}

Giant pulses from the Crab pulsar are long known 
(Staelin \& Reifenstein 1968) but remain enigmatic tools
for probing the pulsar emission mechanism.   Recent work
has established that giant-pulse fluctuations 
are  most likely associated with changes in the coherence 
of the radio emission (Lundgren \etal\ 1995), 
that giant pulses are broadband (Sallmen \etal\ 1999),
and that they are superpositions of extremely narrow
nanosecond-duration structures (Hankins \etal\ 2003). 
Giant pulses from the Crab pulsar have the largest implied
brightness temperature of any known astrophysical source.
A simple estimate for the brightness temperature,
based on the light-travel size and ignoring relativistic dilation, 
is
\be
T_b &=& \frac{S_{\nu}} {2 k}
	\left( \frac{D}{\nu\Delta t}\right)^2
	 \nonumber \\
    &=& 10^{30.1}\,{\rm K}\, 
    	S_{\nu}({\rm Jy})
	\left( \nu_{\rm GHz} \Delta t_{\mu s}\right)^{-2}
	\left (\frac{D}{2\,\rm kpc} \right)^2,
\ee
where 
$S_{\nu}$ is the peak flux density at frequency $\nu$, 
$D$ is the distance, and
$\Delta t$ is the pulse width.   For observed peak amplitudes
and pulse widths (e.g. $10^3$ Jy at 5 GHz with
$\Delta t = 2$ ns; Hankins et al. 2003), 
$T_b$ ranges to as high as $10^{37}$ K. 

In this paper we are concerned with the occurrence of giant pulses
as a function of frequency and also 
where they occur as a function of pulse phase.  We also establish
the properties of the Crab pulsar's emission in order that we
can estimate the detectability of giant pulse emitters from other
galaxies.  Detections of such objects would enable studies
of pulsar populations in those galaxies and use of the pulses to
probe the interstellar medium (ISM) in those galaxies as well
as the intervening intergalactic medium.    In addition,
the Crab pulsar may serve as a prototype of intense, coherent emission
from other classes of high-energy objects which may share
a similar  physical configuration, namely a collimated flow of
relativistic particles.   As such, the Crab pulsar may signify
the presence of other source classes in the transient radio universe
that could be targets for proposed widefield telescopes such as LOFAR
and the SKA.

In \S\ref{sec:obs} we discuss the observations and issues pertaining
to the strong background from the Crab Nebula and 
to the role of scintillation modulations associated with multipath
propagation through the ISM and the Crab Nebula. 
Average profiles and giant-pulse profiles are discussed in
\S\ref{sec:profiles} and timing and amplitude statistics in
\S\ref{sec:timing}.   
Detectability of giant pulses in other galaxies is summarized
in \S\ref{sec:detect} and we conclude the paper in
\S\ref{sec:summary}.   In an Appendix, we discuss frequency
structure caused by intrinsic pulse structure and 
by scintillation. 

\section{Observations}\label{sec:obs}

The Crab pulsar was observed at the Arecibo Observatory in
January to March and May 2002 using receivers in the Gregorian optical
path.      Analog signals were analyzed with a fast-dump, real-time
correlator system, the Wideband Arecibo Pulsar Processor 
(WAPP, \verb+http://www.naic.edu/~wapp+),  which outputs a data stream
of correlation functions at specifiable time intervals.
The number of correlation lags (and hence the number of channels
across a choice of bandwidths after Fourier transformation)
is selectable.
The total bandwidth
used was 100 MHz for all but 0.43 GHz, where we used 12.5 MHz.  
We optimized the time resolution by matching the dump time between correlations
with the dispersion smearing time across individual channels, subject
to a constraint on overall data rate that could be recorded,
$\lesssim 20$ Mbyte s$^{-1}$.
Table~\ref{tab:params}  gives the observing parameters: 
observing frequency, $\nu$ (GHz); 
modified Julian date (MJD); 
total time, $T$ (hr), of acquired data; 
total bandwidth, $B$ (MHz);
sample interval, $\Delta t$ ($\mu s$);
channel bandwidth, $\Delta\nu$ (MHz); 
dispersion smearing across a single channel,
$\Delta t_{\rm DM} = 8.3 \,\mu {\rm s}\, \DM \Delta\nu\ \nu^{-3}$;
and the mean system noise, $\Ssys$, expressed in Jy.
We used a dispersion measure, \DM\ = 56.7910 pc cm$^{-3}$,
to dedisperse the data.

Processing consisted of 
(1) Fourier transforming the  correlation
 	functions for each of two polarization channels; 
(2) summing the resultant spectra for the  two polarizations; 
(3) dedispersing by summing over frequency channels 
	while taking into account time delays associated with
	plasma dispersion in the ISM; 
(4) averaging the time series synchronously with the 
	pulsar period to form a standard intensity profile;
(5) identifying individual giant pulses and their occurrence
	times by selecting intensity samples that exceeded
	the off-pulse mean by 5$\sigma$;
(6) synchronously averaging the individual giant pulses to
	form a histogram of giant pulses vs. pulse phase;
and
(7) aligning average profiles and individual giant pulse
	profiles in pulse phase by using TEMPO and a spin model
	for the Crab pulsar.  We also used TEMPO to perform
	an arrival-time analysis on individual giant pulses,
	as discussed in \S\ref{sec:timing}.
	For the TEMPO analysis, we used timing models from the
	Jodrell Bank timing program 
	(\verb+http://www.jb.man.ac.uk/~pulsar/crab.html+).

The procedure for finding giant pulses followed that of
Cordes \& McLaughlin (2003).   The dedispersed time series 
was first analyzed with the original time resolution and
then progressively smoothed and decimated by factors of
two in order to approximately match filter to pulses with different widths.
In the end, most pulses were optimally detected with no smoothing or
only one level of smoothing, as is consistent with the known properties
of giant pulses and average profiles 
(Moffett \& Hankins 1996; Sallmen et al. 1999).

\subsection{Importance of Nebular Background}


The Crab Nebula, whose flux density $\sim 955\nu^{-0.27}$ Jy
($\nu$ in GHz)
(Allen 1973; Bietenholz \etal\ 1997), dominates
the system temperature if it is not resolved by the telescope.
Define the system temperature in the absence of the Crab Nebula as
$\Tsyso$ and the contribution from the Crab Nebula as $\TCN$.  Expressing
these in flux density units by dividing by the telescope
`gain' G (K Jy$^{-1}$), the total system noise level is
\be
\Ssys = \Ssyso + \SCN.
\ee
For the Crab pulsar and Nebula, the system temperature is strongly influenced
by the Nebula if $\SCN > \epsilon\Ssyso$ with $\epsilon = 0.1$, say.  
For a single-dish telescope with
typical system temperature $\Ssyso = 50$ K and 60\% aperture efficiency,   
this condition is satisfied for antenna diameters that satisfy
\be
d > 17 \,{\rm m}\,\, \nu^{0.13}\left ( \frac{\epsilon\Tsyso}{50\, K} \right)^{1/2}. 
\ee
For very large telecopes (either large single-dish antennas or
arrays),   the effective beam width can be smaller than the Crab Nebula,
reducing its contribution to the system noise by a factor
$f_{\nu} = \Omega_{\rm A} / \Omega_{\rm CN}$,  
where $\Omega_{\rm A}$ is the solid angle of the primary antenna beam,
and $\Omega_{\rm CN}$ is the solid angle of the Crab Nebula. 
The total system noise level is then
\be
\Ssys = \Ssyso + f_{\nu} \SCN,
\ee
The characteristic diameter (geometric mean of the major and minor
axes) of the Crab Nebula is approximately 5.5 arcmin.  
For the Arecibo telescope, our data at 
0.43 GHz, with beam size equal to 11 arcmin (FWHM)
include all of the flux density of the Crab Nebula,
while higher frequencies resolve out some of the flux density.
The last column of Table~\ref{tab:params} indicates our estimate of
the system noise including any dilution of the Crab Nebula's contribution.

In the following, we express pulse amplitudes in terms of the mean system
noise.

\subsection{Scintillations}
\label{sec:scints1}

Diffractive interstellar scintillation (DISS) strongly influences
the detectability of the pulsar at frequencies of 3 GHz
and higher.  There are three regimes that may be identified
for scintillation modulations, depending on the size
of the scintillation bandwidth, $\dnud$, relative to the
total bandwidth $B$ and to the channel bandwidth, $\Delta\nu$.
For $\dnud \ll \Delta\nu$,  scintillations are essentially
quenched because individual `scintles' are averaged out.
For $\Delta\nu \lesssim \dnud \lesssim 0.2 B$,   scintillations are
identifiable as frequency structure in the spectrum of a 
strong, individual pulse; the net modulation in the dedispersed
time series depends on the number of scintles across $B$ and
conforms roughly to a $\chi^2$ distribution with
$\sim 0.4B/\dnud$ degrees of freedom.   Finally, for
$\dnud \gtrsim 0.2B$,   the dedispersed time series is fully
modulated by DISS, with an amplitude modulation factor conforming
to a one-sided exponential.    The factor, $0.2 B$, represents
the approximate value of $\dnud$ for which we would expect
only one scintle within the total bandwidth, $B$. Taking into account that
$\dnud \propto \nu^{4.4}$, DISS becomes progressively
more important in going to higher frequencies until the scattering becomes
weak (in the sense of phase perturbations on the Fresnel scale
becoming less than 1 radian; see, e.g., Rickett 1990).

We estimate the DISS bandwidth by using the relation
$2\pi\dnud\taud = C_1$ (Cordes \& Rickett 1998; hereafter CR98), where
$\taud$ is the pulse-broadening time and $C_1$ is
a constant dependent on the spectrum and spatial distribution
of scattering irregularities; we adopt $C_1 = 1.05$, a value 
appropriate for a thin screen.
The pulse-broadening time for the Crab pulsar
is known to vary 
(e.g. Isaacman \& Rankin 1975; Lyne \& Thorne 1975; Backer \etal\ 1998; 
Lyne, Pritchard \& Graham-Smith 2001; Backer \etal\ 2000), ranging
from about 0.28 ms to 1.3 ms at 0.3 GHz (Sallmen \etal\ 1999).
Adopting $\taud(0.3 \,{\rm GHz}) = 0.5$ ms as a reference value,
we estimate 
$\dnud \approx 67 \,{\rm kHz}\, 
\nu^{4.4} [0.5\,{\rm ms}/\taud(0.3\,{\rm GHz})]$, 
with $\nu$ in GHz. 
Using this expression, we expect that the dedispersed pulse
will show fully modulated 
scintillations (i.e. after summing over the 100 MHz bandwidth)
for $\nu \gtrsim 3.6$ GHz.   

We found the mean pulsar flux density (averaged over a few
minutes) to be heavily modulated on time scales
as short as 5 minutes at frequencies $\gtrsim 3$ GHz and nearly
unchanging at lower frequencies.  Such fluctuations are consistent
with those expected from DISS,  as we discuss
in \S\ref{sec:scints2}.  We also saw epoch-to-epoch
fluctuations (time scales of one day and longer) that are
consistent with refractive interstellar scintillations (RISS),
like those identified by Lundgren \etal\ (1995) at 0.8 GHz
on time scales of a few days.  The combination of RISS and DISS
is particularly strong at 8.8 GHz where the pulsar is
undetectable on many days but quite bright, in the mean,
on occasional days with DISS fluctuations contributing on 
shorter time scales.  We note that at 8.8 GHz, the diffraction
bandwidth $\dnud \approx 1$ GHz (assuming the $\nu^{4.4}$
scaling), implying that the strength of scattering (as
defined in scintillation literature; see Rickett 1990) is
not strong and that the DISS and RISS `branches' are not
as distinct as at lower frequencies (e.g. Narayan 1992).

\section{Average Profiles}\label{sec:profiles}

We calculated average profiles of the total intensity
by summing partial-sum profiles (of 3 to 12 minutes duration,
depending on frequency) 
in which the largest pulse component had a signal-to-noise
ratio (S/N) larger than five and which were unmarred
by radio frequency interference (RFI).   
At frequencies higher than 0.43 GHz,  typically  only a minority of
profiles were included owing to the effects of DISS or RFI.
Likewise, profiles of giant pulse
counts were calculated using data subsets
corresponding to those included in the average intensity
profiles.

\begin{figure*}
\epsfxsize=\hsize
\epsfbox{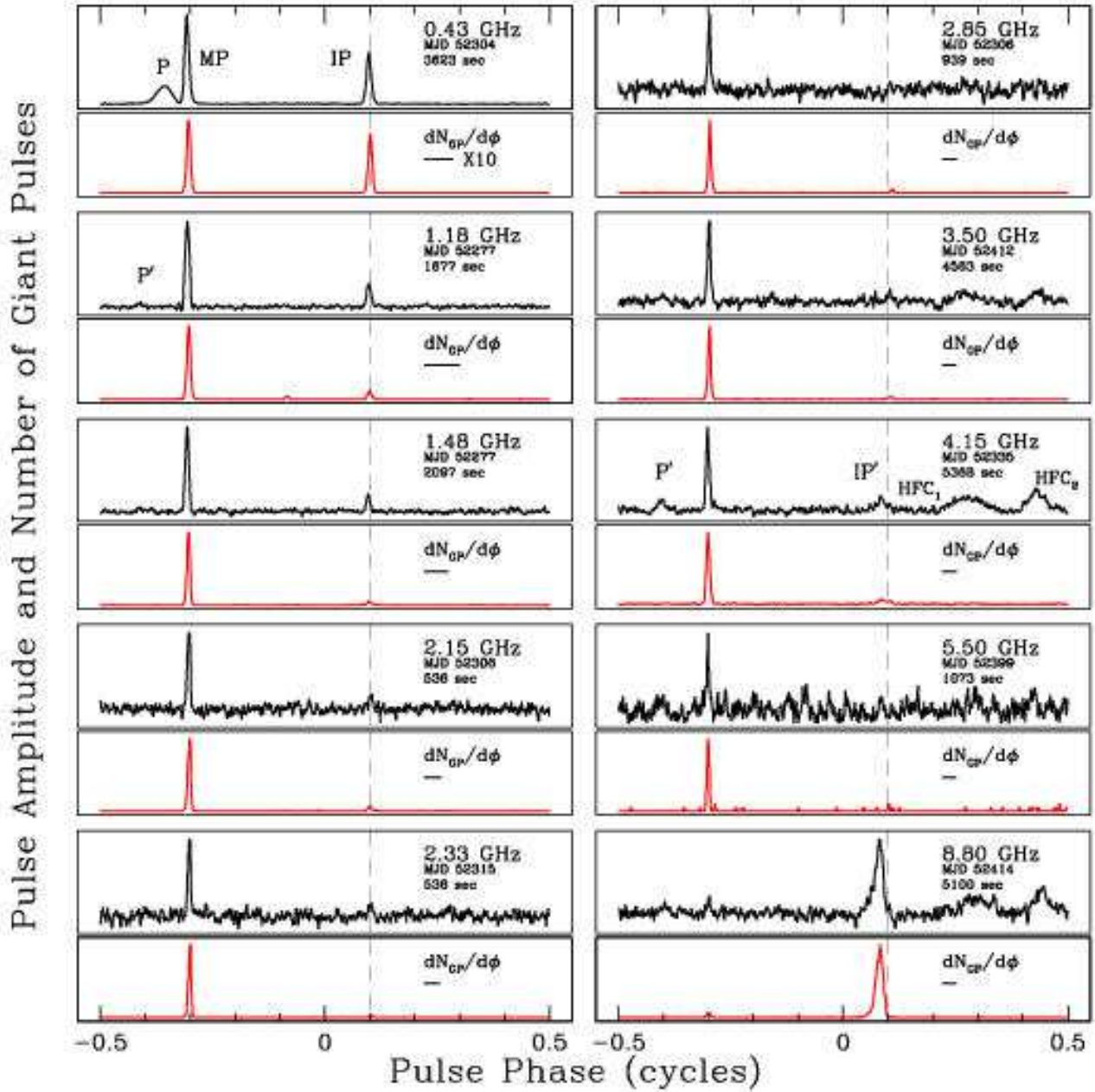}
\figcaption{
Total intensity profiles of the Crab pulsar at 10 radio frequencies.
The pair of plots for each frequency is the standard average intensity
profile (top) and a histogram of counts of giant pulses plotted against
pulse phase (bottom).   A threshold of $5\sigma$ was used to obtain 
giant pulses included in the histograms.
The total integration time is given in the top
frame and a horizontal bar designates the net instrumental
time resolution, including the effects of dispersion smearing
across individual spectrometer channels;  the shown
bar length is ten times the actual length.
\label{fig:composite}
}
\end{figure*}

Figure~\ref{fig:composite} shows average intensity and giant-pulse
profiles for the ten frequencies journaled in 
Table~\ref{tab:params}.    The top panel of the pair shown
for each frequency is the total intensity profile and the bottom
panel shows the giant-pulse histogram vs. pulse phase,
i.e. the number of giant pulses in the given pulse phase bin
that are above threshold ($5\sigma$).    In some of the panels
we designate features in the pulse profile, including
low-frequency precursor pulse (P), 
main pulse (MP),
interpulse (IP),
an intermediate-frequency precursor component 
	(P$^{\prime}$; referred to by Moffett \& Hankins (1996) as
	a `low-frequency component');
a shifted interpulse component appearing at mid-to-high
	frequencies (IP$^{\prime}$);
and two high-frequency components (HFC1 and HFC2) that were first
identified by Moffett \& Hankins (1996).

\smallskip

We point out the following features of the set of profiles:
\begin{enumerate}
\item The pulsar is more readily detectable in its single
giant pulses than in the average pulse  at all frequencies.
This is manifest by the larger S/N in the histogram plots
compared to the average profile plots in Figure~\ref{fig:composite}.
\item There is strong evolution of the relative strength of
	MP and IP as a function of frequency.   The ratio
	of peak IP to peak MP steadily declines from 0.43 to
	2 GHz,  stays low from 2.5 to 3.5 GHz,   
	and rises at higher frequencies so that the IP is much
	stronger than the MP at 8.8 GHz.
\item At 4.15 GHz, IP$^{\prime}$ appears approximately
	0.03 cycles before the location of the lower frequency IP
	and becomes very strong relative to the MP at 8.8 GHz.
\item At frequencies of 3.5 GHz and higher two new components,
	HFC1 and HFC2, appear and persist up to the highest frequency
	we used (8.8 GHz).   

\item It is unclear if the HFC1 and HFC2 components are present 
	at 2.15, 2.33 and 2.85 GHz owing to the low
	S/N of those profiles, which derives from the short integration
	times and the effects of scintillations.
\item Giant pulses occur only in MP, IP and IP$^{\prime}$.
	For this reason we conclude that IP and IP$^{\prime}$ are
	probably associated with the same physical emitting 
	region or beam in the pulsar. 
\item There is more scatter in the pulse phase of the interpulse
	at high frequencies, manifested in the broader width of the
	interpulse component.   As discussed below in \S\ref{sec:timing},
	at 8.8 GHz the phase residuals
	appear to show a two-component distribution, one centered
	on $\phi = 0.38$ cycles, the other representing giant pulses
	skewed toward smaller phases.
\end{enumerate}

Table~\ref{tab:gpstats} gives 
the frequency and total time span used in our analysis in columns 1 and 2;
column 3 gives the mean mainpulse-to-interpulse pulse-phase difference; 
column 4, the ratio of the numbers of detected interpulses and mainpulses;
and the last
column gives the detection rate for events found at $8\sigma$ or 
higher.  At 0.43 GHz, about 1 in 10 pulses is detected above this
level.  We emphasize that the detection rate is epoch dependent
owing to scintillation modulations.   Slow, refractive scintillations
affect the rates at all frequencies, while fast diffractive scintillations
are particularly important at frequencies above 3 GHz.  The 8.8 GHz
results apply to a particular day when the scintillation modulation
boosted intensities far above their normal level.
The mainpulse to interpulse phase difference varies systematically with
frequency, remaining constant with frequency (within errors) from
0.43 to 3.5 GHz and then declining with increasing frequency up
to 8.8 GHz.  This variation in calculated phase difference corresponds
to the shift from IP to \IPprime.

\section{Amplitude and Timing Statistics of Giant Pulses}\label{sec:timing}

Figures ~\ref{fig:snrhist430MHz} and \ref{fig:snrhist8.8GHz} show
histograms of signal-to-noise ratio (S/N) at 0.43 and 8.8 GHz 
for the main and interpulse components specified in 
Figure~\ref{fig:composite}.  
We include only pulses with $S/N > 8$ in order to provide consistency
with the timing analysis discussed below, where we use an 8$\sigma$
threshold for the purpose of obtaining timing residuals minimally
influenced by noise.
Traditionally, giant-pulse
amplitude distributions have been characterized as power laws
(e.g. Argyle \& Gower 1972; Lundgren \etal\ 1995).   The histograms
shown here have roughly power-law segments
to their distributions but there 
are outlier pulses at especially high S/N 
at both frequencies.  Roughly, a power law with slope 
$\approx -2.3$ can be drawn through the MP histogram at 0.43 GHz
in Figure ~\ref{fig:snrhist430MHz} and a slope
$\approx -2.9$ at 8.8 GHz (Figure~\ref{fig:snrhist8.8GHz}.
These can be compared with slopes of approximately -2.5 at 0.146 GHz
(Argyle \& Gower 1972) and -3.6 at 0.812 GHz (Lundgren \etal\ 1995).  Overall
there thus appears to be steepening of the histogram in going
from low to high frequencies.
Remarkably, the largest pulse
at 0.43 GHz has S/N $\sim 1.1\times 10^4$, which  
is inconsistent with the probability
implied by the power law at lower S/N.  We suggest
that this pulse is an example of a
{\it supergiant} pulse.   The same is true at
8.8 GHz,  where giant pulses in the interpulse region
are dominant in number but the largest pulse 
appears in the main pulse component and is 
10$\times$ larger than the largest interpulse
giant pulse.    Conceivably, the largest pulses
seen at 0.43 and 8.8 GHz are statistical flukes.
The extensive observations of Lundgren et al. (1995)
at 0.8 GHz do not indicate the  presence of a gap between
the  brightest and typical giant pulses.

\smallskip
\begin{center}
\epsfxsize=\hsize
\epsfbox{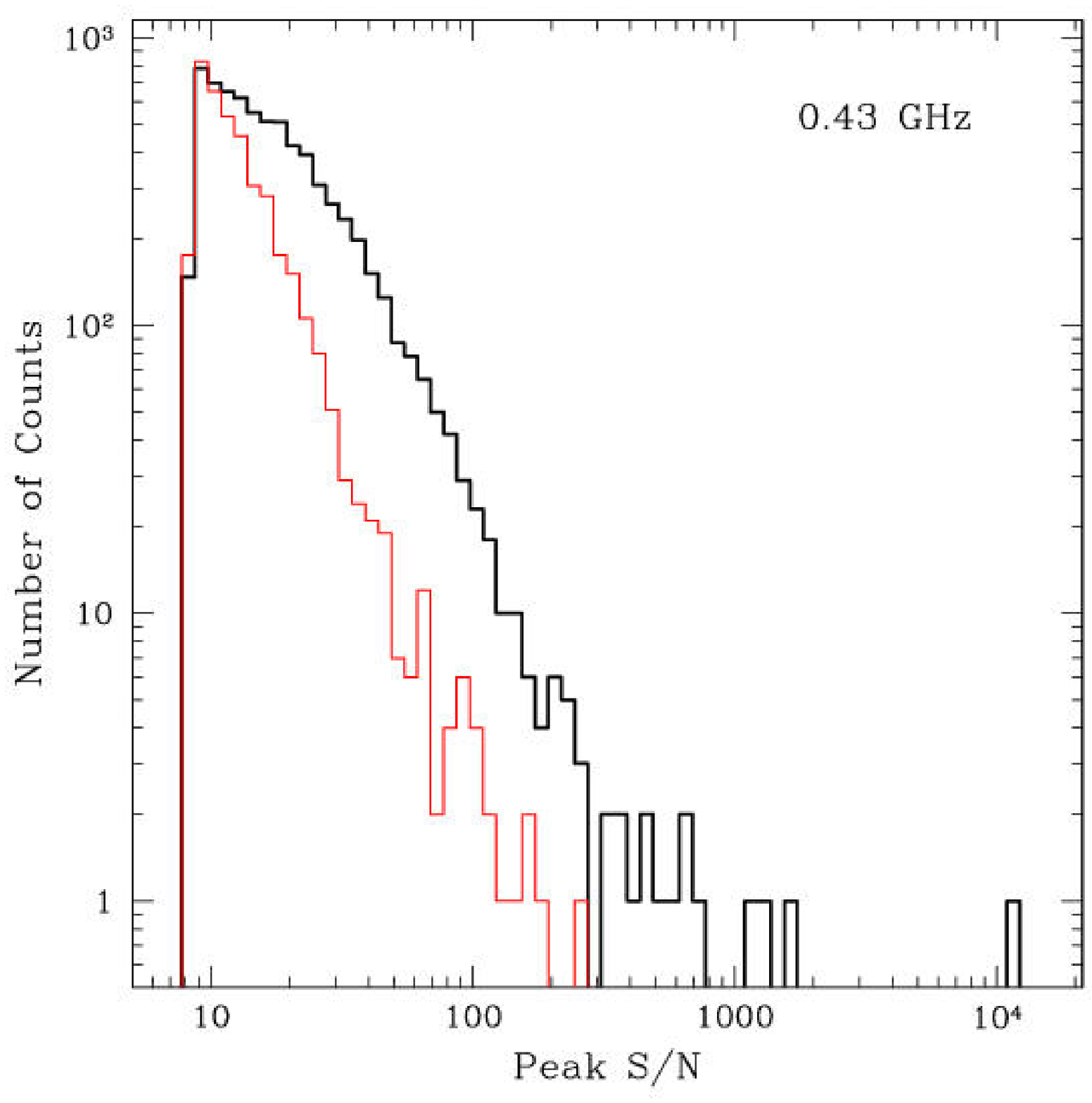}
\figcaption{
\label{fig:snrhist430MHz}
Histogram of giant-pulse peak amplitudes at 0.43 GHz. The red curve is
for the interpulse and the black curve for the main pulse.
}
\end{center}

\smallskip
\begin{center}
\epsfxsize=\hsize
\epsfbox{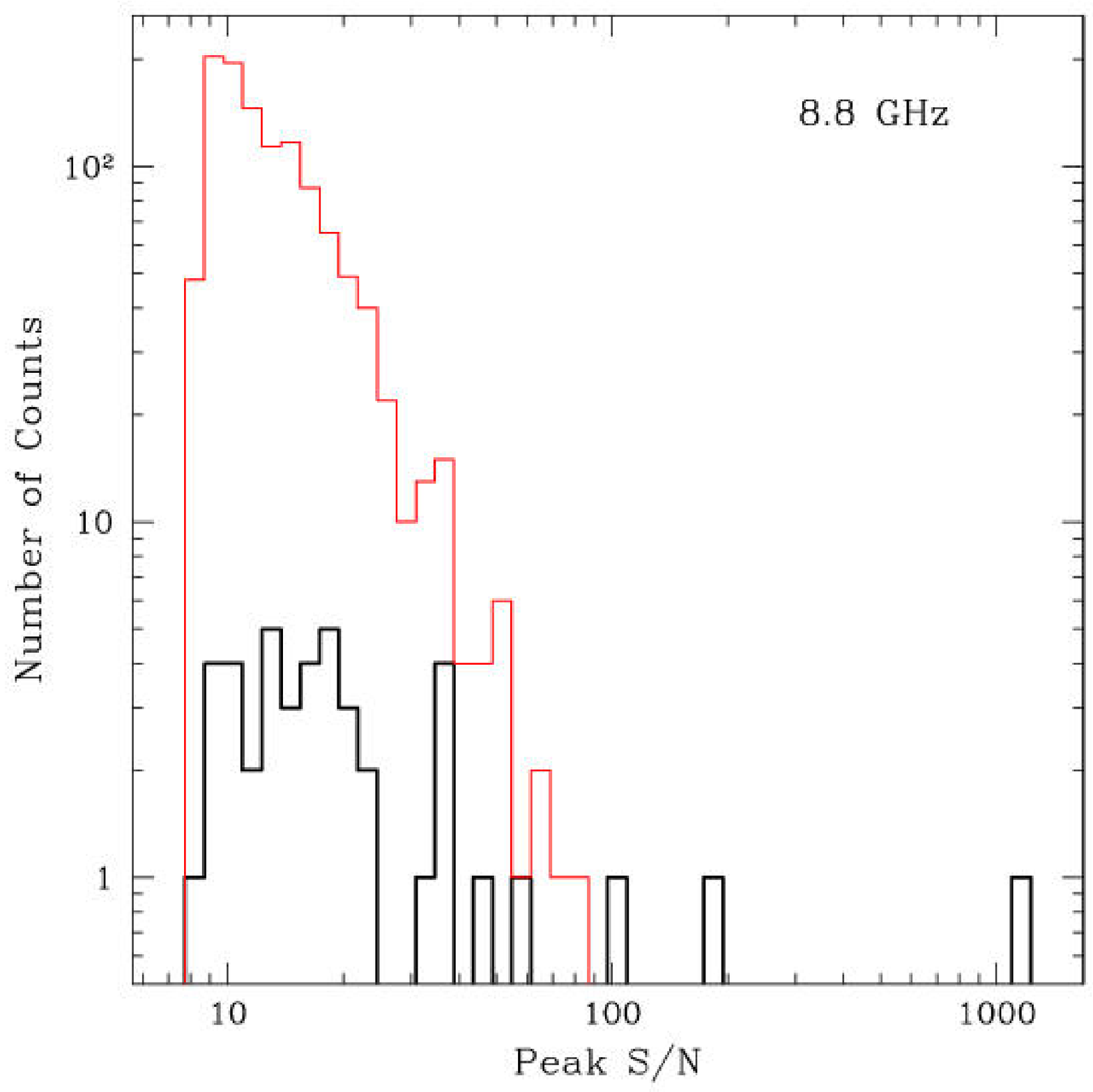}
\figcaption{
\label{fig:snrhist8.8GHz}
Histogram of giant-pulse peak amplitudes at 8.8 GHz. The red curve is
for the interpulse and the black curve for the main pulse.
}
\end{center}

The joint statistics of timing phase residuals and pulse amplitudes
(expressed as S/N) are shown in 
Figures ~\ref{fig:phivamp430MHz} and ~\ref{fig:phivamp8.8GHz}
for the mainpulse and interpulse separately.  
At the lower frequency (0.43 GHz),
the mainpulse phase residuals show a skewed distribution toward
larger phases.  At 8.8 GHz, the distribution of
phase residuals in the interpulse is much broader than in the mainpulse
and in either component at 0.43 GHz.   This trend is consistent with
the appearance of the average profiles in Figure ~\ref{fig:composite}.
At 8.8 GHz,  the giant interpulses
showing the most negative phase residuals tend to be weaker than the
average.   Otherwise, there is no evidence for a strong relationship
between amplitude and phase residual. 

\smallskip
\begin{center}
\epsfxsize=\hsize
\epsfbox{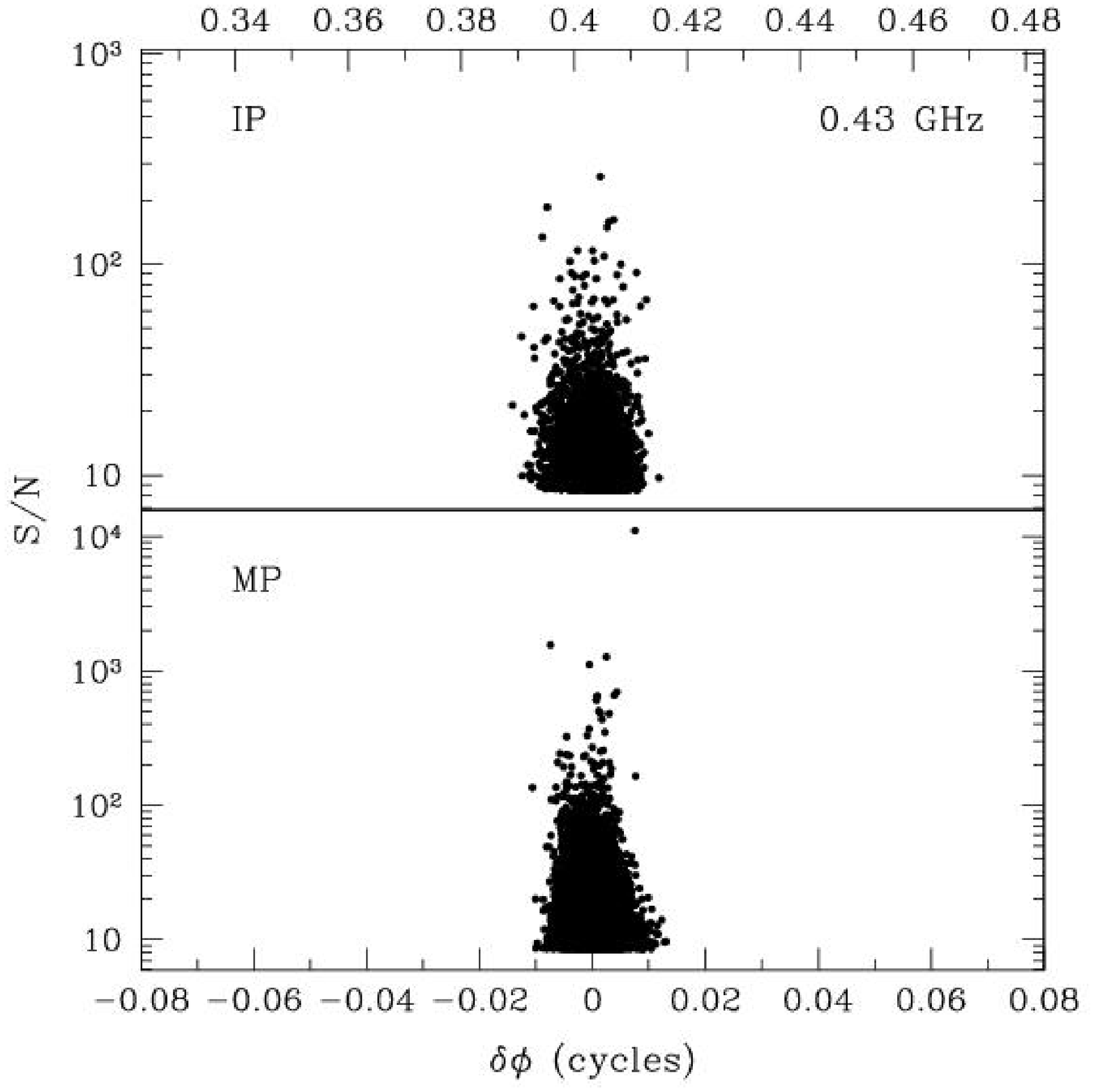}
\figcaption{
\label{fig:phivamp430MHz}
Scatter plots of signal-to-noise ratio and pulse phase 
for 0.43 MHz for the interpulse (\IPprime) and mainpulse (MP).
The mean mainpulse phase is defined to be zero.
}
\end{center}

\smallskip
\begin{center}
\epsfxsize=\hsize
\epsfbox{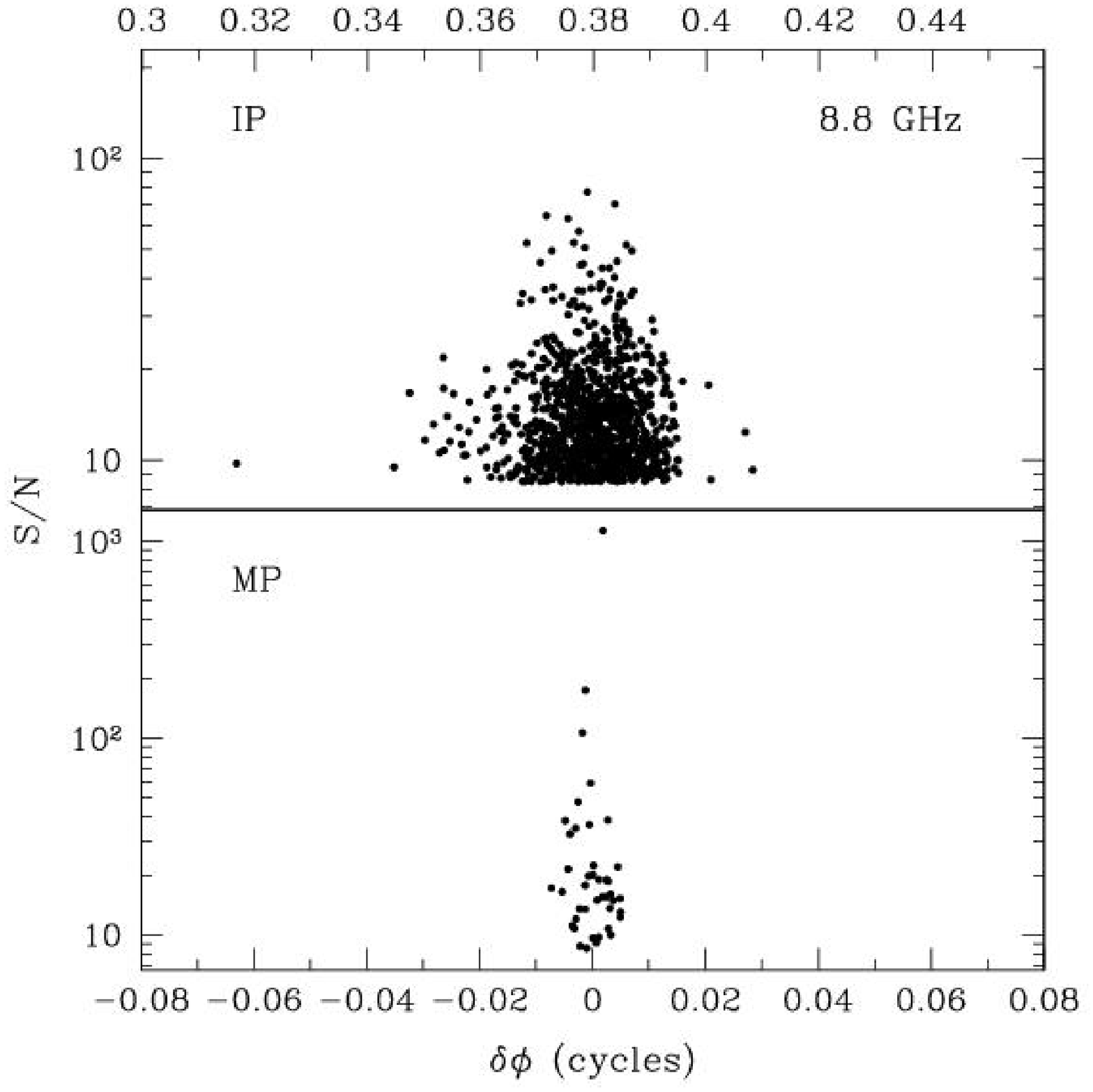}
\figcaption{
\label{fig:phivamp8.8GHz}
Scatter plots of signal to noise ratio and pulse phase residual
for 8.8 GHz for the interpulse (IP) and mainpulse (MP).
}
\end{center}

\section{Scintillations and Spectral Fluctuations}\label{sec:scints2}


From the discussion in \S\ref{sec:scints1}, we estimate that
the scintillation bandwidth
will range from about 2 kHz at 0.43 GHz to about 1 GHz at 8.8 GHz
if we assume a scaling $\dnud \propto \nu^{4.4}$.  
Previous work also suggests that at any epoch the actual
scintillation bandwidth could vary by a factor of a few about these
values.  Such variations are due to the stochastic nature of
the process but also are caused by refractive modification
of the diffraction parameters (e.g. Rickett 1990).
At our lowest
frequency, 0.43 GHz, the DISS bandwidth is about 1/10 the channel
bandwidth (c.f. Table~\ref{tab:params}).   Thus, the brightest features
in the frequency-time data are expected to be diminished 
by the smoothing implied by  the spectrometer resolution.    At our 
highest frequency, 8.8 GHz,  the predicted DISS bandwidth is a
factor of eight smaller than the center frequency, 
signifying that scintillations are
in the transition regime between strong and weak scattering.
In the transition regime, we expect deep modulations as in the
strong scattering regime, but with different statistics
and, according to our estimate, they will be highly correlated
across our 100 MHz bandwidth.  

Of course, our predictions
for DISS bandwidth require some caveats.  First, our data were
obtained over a 0.4 year period over which time the scattering
strength undoubtedly varied, probably yielding an implied
pulse-broadening time at 0.43 GHz different from the value
we have assumed ($100\,\mu s$).  Second, the scaling with frequency
of the DISS quantities may depart from that which has been
identified along other lines of sight in the ISM.   The scaling
with the 4.4 exponent has been established for pulsars with small
DMs, while a few objects show a weaker scaling as $\nu^4$.  Recent
work on high-DM pulsars (L\"ohmer \etal\ 2001; Bhat \etal\ 2003)
indicates that the pulse broadening time may vary as weakly
as $\nu^{-3}$ (and the scintillation bandwidth thus as
$\nu^{3}$).  However, it is also clear from Bhat \etal\ (2003)
that empirical determinations of the exponent are sensitive to 
assumptions about the form of the pulse-broadening function fitted
to the data and how it interacts with the assumed intrinsic
pulse shape of the pulsar.    Nonetheless, despite these uncertainties,
evidence suggests that the pulse broadening from the Crab pulsar is
not only highly variable, but often exceeds the predictions based on
lower frequencies using either of the strong scaling laws,
$\dnud\propto\nu^{4}$ or $\nu^{4.4}$.    Possible interpretations
include contributions from scattering within the pulsar magnetosphere
(Hankins \& Moffett 1998)
or from scattering regions within the Crab Nebula that are bounded spatially
(Cordes \& Lazio 2001).   Spatially bounded scattering regions
can generate scattering times that scale with frequency differently
than with an exponent of 4 or 4.4.

In Figures ~\ref{fig:tfplot.430MHz} - \ref{fig:tfplot.8.8GHz} we show
plots of the pulsed flux for single giant pulses 
in the frequency-time plane, the pulse
shape obtained by summing over frequency both with and without
compensation for dispersion delays, and the spectrum of the pulse.

At 0.43 GHz,  the pulse is easily detected even without dedispersion
owing to the high S/N.
Structure in the spectrum is quite spiky and is 
associated with individual scintles caused by DISS.   This is so,
in spite of the fact that the DISS bandwidth is substantially smaller than
the channel bandwidth, because the spacing between scintles is quite
large.   DISS in the strong scattering regime is exponentially
distributed and with an ensemble average mean modulation of unity.
Consistent with the statistics is the estimate that the
number of strong scintles within a bandwidth $B$ is 
\be
N_{\nu} \approx 1 + 0.2 B / \dnud.
\ee
As can be seen  in the right-hand panel of Fig~\ref{fig:tfplot.430MHz},
the minimum spectral values away from the bandpass edges
are well offset from zero, signifying that 
the overall modulation is less than the 100\%
expected from exponential DISS statistics, 
consistent with the smoothing of scintles that occurs in the 
process of channelizing the data. 

By comparison, plots of giant pulses  at 1.5 and 2.4 GHz 
(Figures~\ref{fig:tfplot.1.475GHz} and \ref{fig:tfplot.2.33GHz})
show minimum spectral values nearly equal to zero flux density,
consistent with the larger, nearly-resolved or resolved 
scintillation structure expected
at those frequencies.    At 2.85 GHz,  the minimum spectral
values are well above zero, signifying that the DISS bandwidth
is large enough that only one or two scintles are expected
across the band.
At 8.8 GHz (Figure~\ref{fig:tfplot.8.8GHz}), 
the modulation of the flux across the bandpass has a much different
character, as expected if $\dnud \gg B$, where $B = 100$ MHz.

\subsection{Scintillation Bandwidths}

We estimate the scintillation bandwidth by calculating
the 
intensity autocorrelation function,
$A(\delta\nu) = \langle I(t,\nu) I(t, \nu+\delta\nu)\rangle$
for the spectrum of each giant pulse and summing over giant pulses.
For this analysis, we used giant pulses with
signal to noise ratios S/N $> 20$
in the dedispersed pulse. 
Scintillation structure is unresolved at frequencies below 2 GHz
and is comparable to or larger than our receiver bandpass at
frequencies larger than 4 GHz.  
Results are shown in Table~\ref{tab:scints} along with scintillation
time scales, discussed in the next section, and the number of giant
pulses used to estimate the parameters.


\smallskip
\begin{center}
\epsfxsize=\hsize
\epsfbox{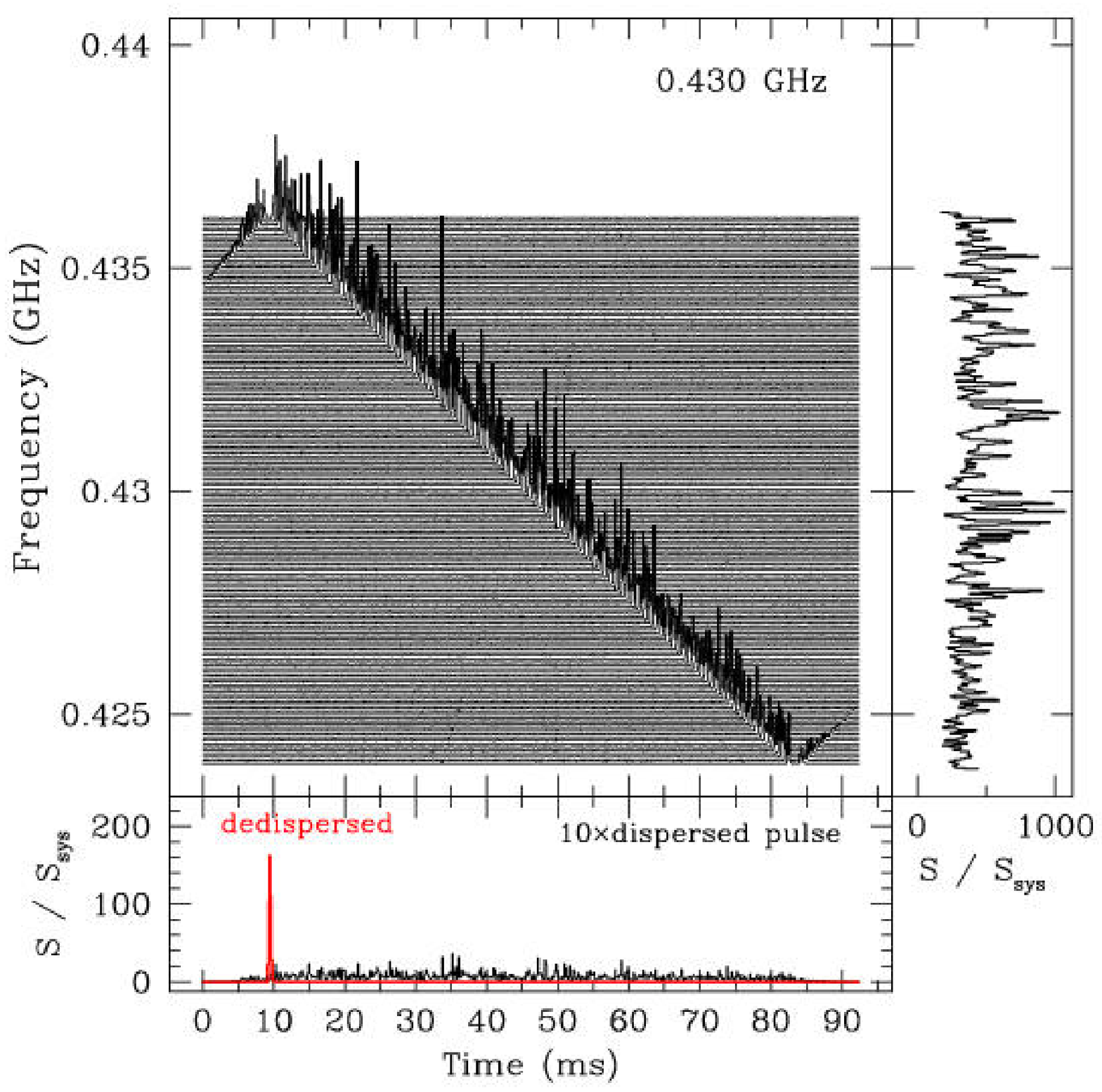}
\figcaption{
\label{fig:tfplot.430MHz}
Plot of intensity against time and frequency, showing a single
dispersed pulse as it arrives at different frequencies centered
on 0.43 GHz.
The right-hand panel shows the pulse amplitude vs. frequency
while the bottom panel shows the pulse shape with and without
compensating for dispersion delays.  This pulse is the largest
in one hour of data, has S/N $\sim 1.1\times 10^4$, and
a pulse peak that is $130$ times the flux density of the Crab Nebula,
or $\sim 155$ kJy.  Note that the segments at either end of the
bandpass where the pulse arrival time is opposite the trend
at most frequencies is caused by aliasing of the signal.
}
\end{center}

\smallskip
\begin{center}
\epsfxsize=\hsize
\epsfbox{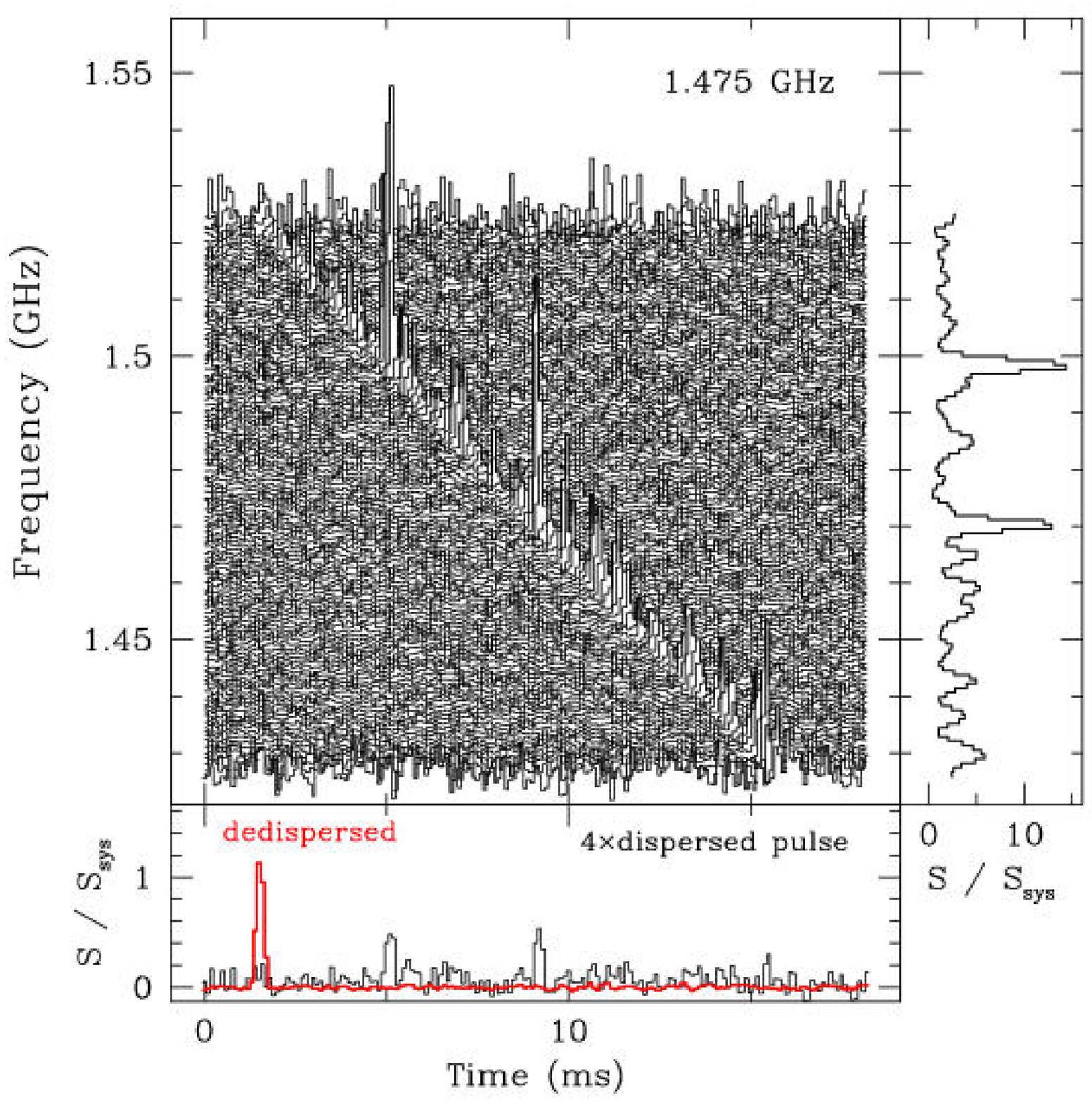}
\figcaption{
\label{fig:tfplot.1.475GHz}
Same as Fig.~\ref{fig:tfplot.430MHz} for a single pulse at 1.475 GHz.
This pulse is the largest
in one hour of data, has S/N $\sim 225$, and
a pulse peak that is $1.2$ times the mean system noise,
or $\sim 1.03$ kJy.  Note that individual `scintles' in the spectrum
reach 14 times $\Ssys$, or 4.1 kJy.
}
\end{center}

\smallskip
\begin{center}
\epsfxsize=\hsize
\epsfbox{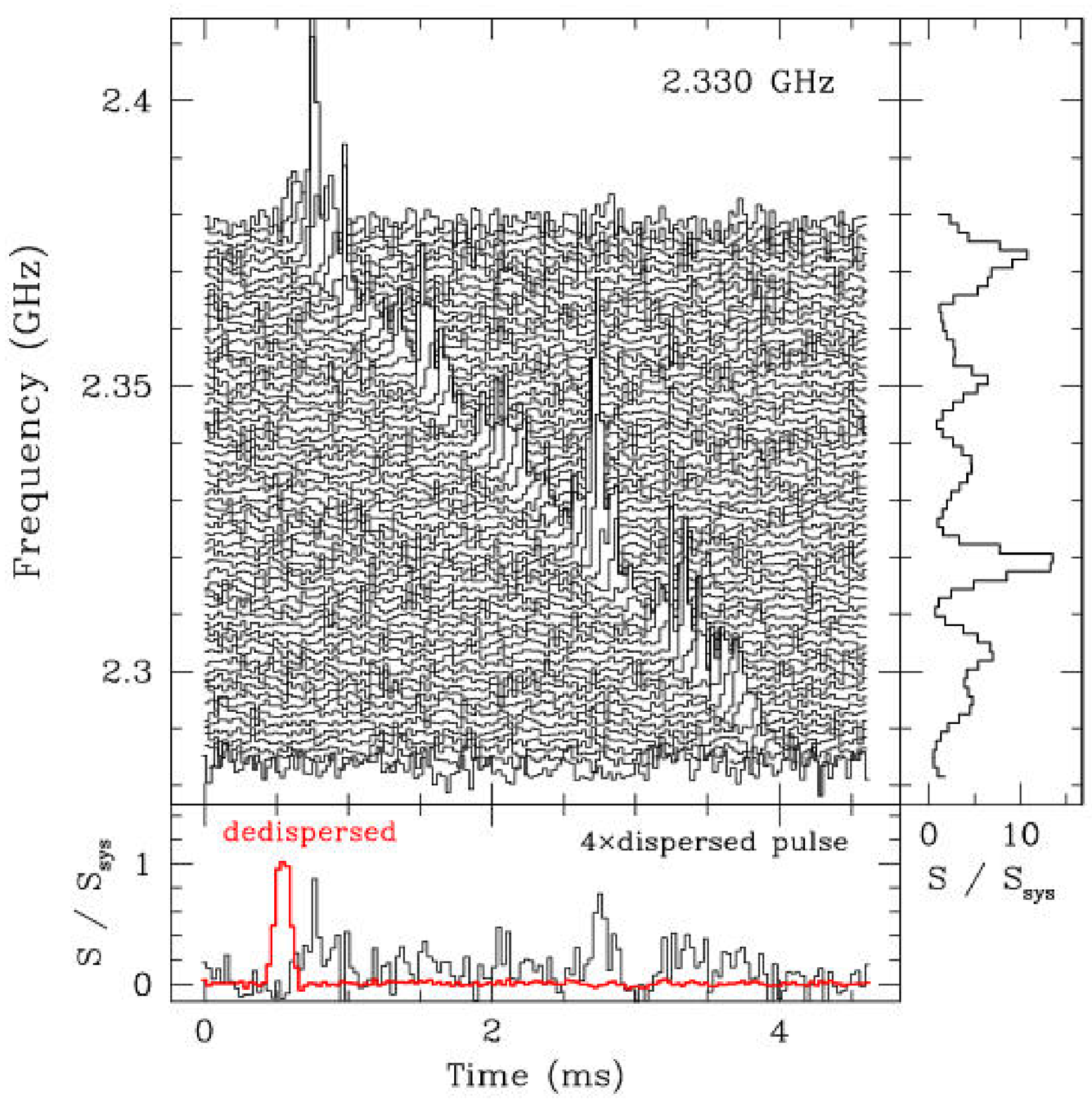}
\figcaption{
\label{fig:tfplot.2.33GHz}
Same as Fig.~\ref{fig:tfplot.430MHz} for a single pulse at 2.33 GHz.
This pulse is the largest
in one hour of data, has S/N $\sim 161$, 
a pulse peak that is $1.1\Ssys$, and a scintillation
peak $\sim 12 \Ssys$.  
At this frequency the telescope beam
resolves the Crab Nebula,  so the peak flux densities are
$\sim 86$ and $\sim 936$ Jy in the dedispersed pulse and
spectrum, respectively.
}
\end{center}

\smallskip
\begin{center}
\epsfxsize=\hsize
\epsfbox{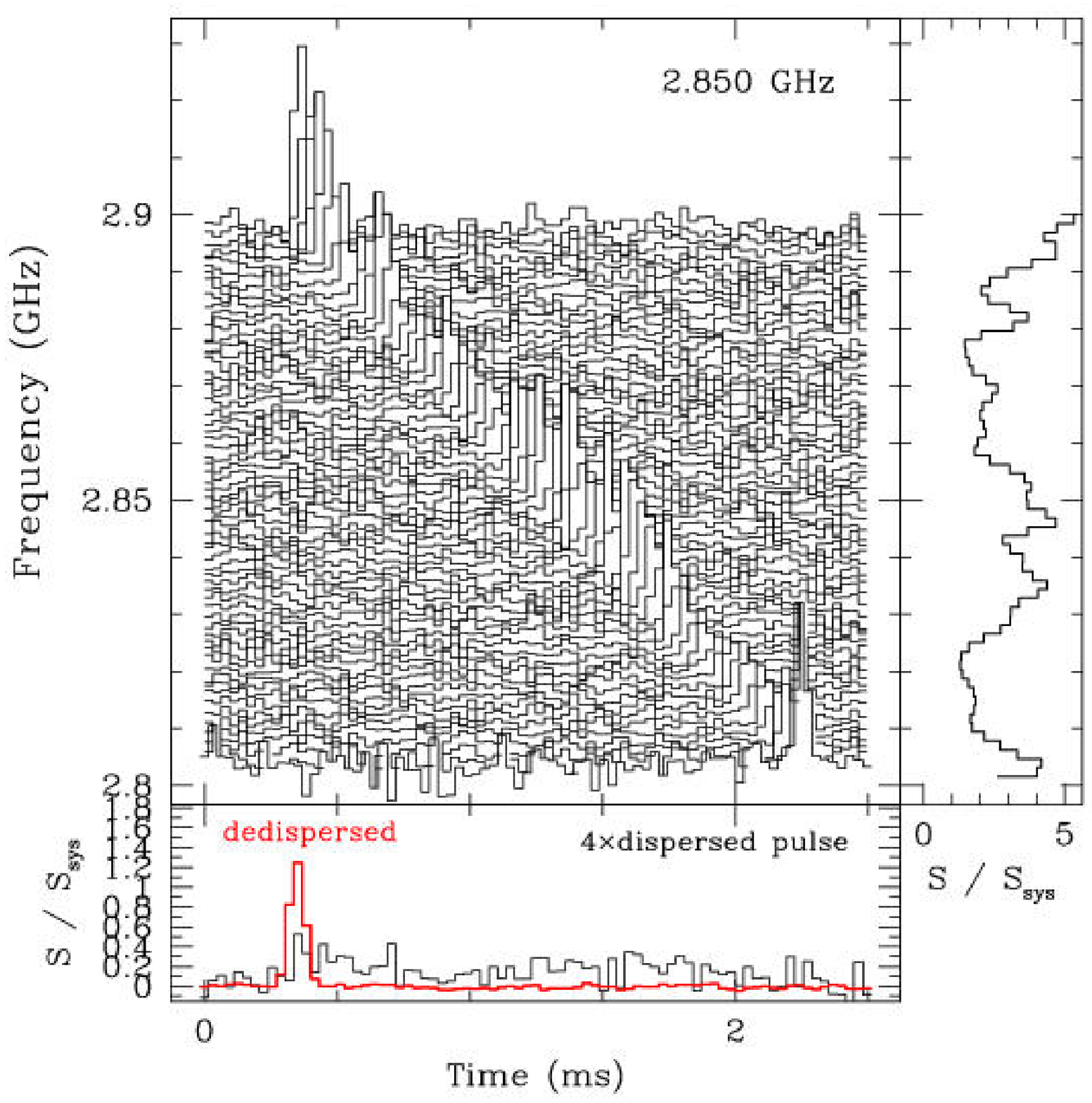}
\figcaption{
\label{fig:tfplot.2.85GHz}
Same as Fig.~\ref{fig:tfplot.430MHz} for a single pulse at 2.85 GHz.
This pulse is the largest
in one hour of data, has S/N $\sim 111$, 
a pulse peak that is $1.2\Ssys$, and a scintillation
peak $\sim 4.5 \Ssys$.  
At this frequency the telescope beam
resolves the Crab Nebula,  so the peak flux densities are
$\sim 89$ and $\sim 333$ Jy in the dedispersed pulse and
spectrum, respectively.
}
\end{center}

\smallskip
\begin{center}
\epsfxsize=\hsize
\epsfbox{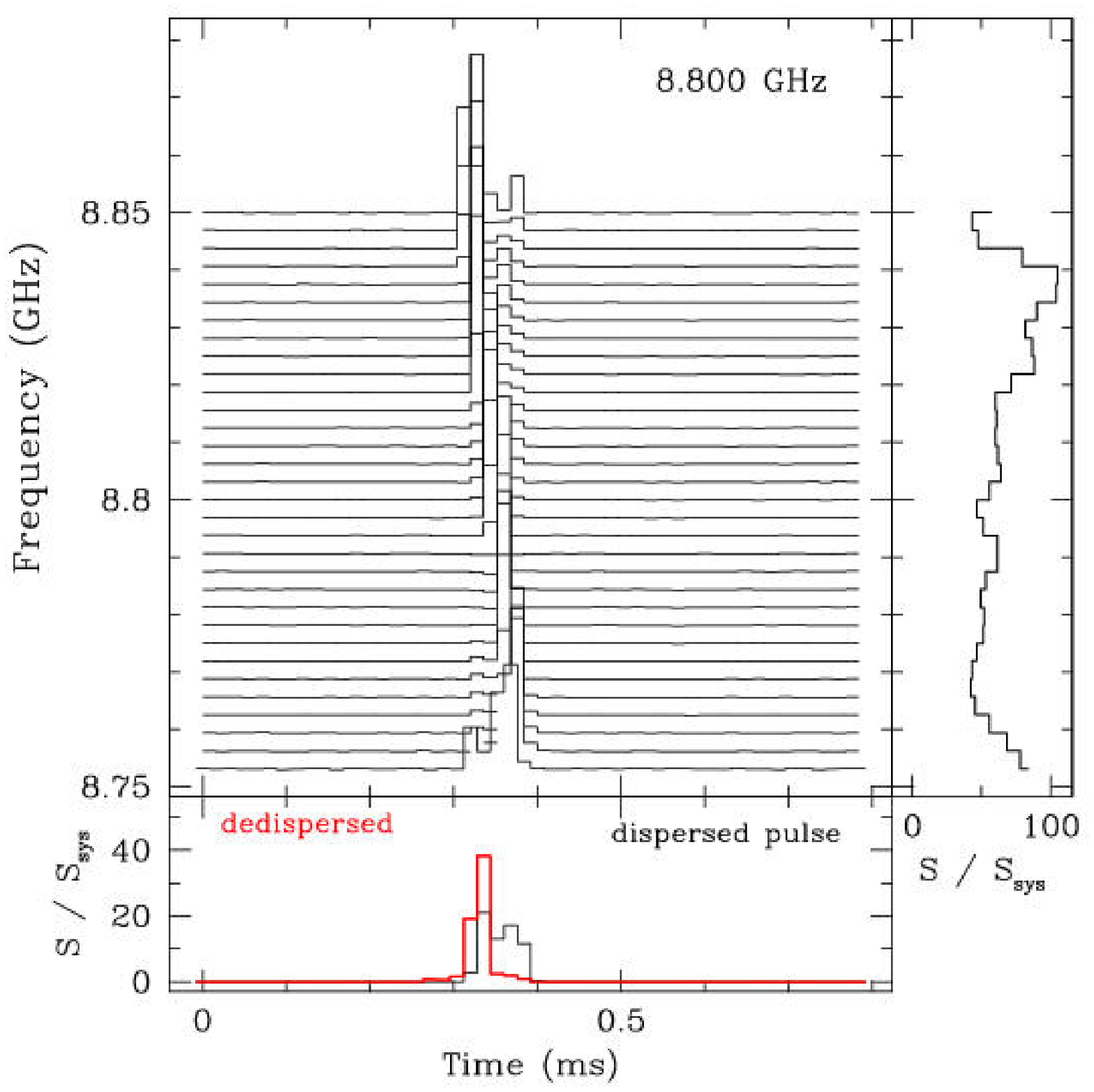}
\figcaption{
\label{fig:tfplot.8.8GHz}
Plot of intensity against time and frequency, showing a single
dispersed pulse as it arrives at different frequencies centered
on 8.8  GHz.
The right-hand panel shows the pulse amplitude vs. frequency
while the bottom panel shows the pulse shape with and without
compensating for dispersion delays.  This pulse is the largest
in one hour of data, has S/N $\sim 1.3\times 10^3$, 
a pulse peak that is $\sim 40\Ssys$, and a spectral peak
$\sim 100\Ssys$.  
At this frequency the telescope beam
resolves the Crab Nebula.  The peak flux densities are
$\sim 880$ and $\sim 2.2\times 10^3$ Jy in the dedispersed pulse and
spectrum, respectively.   
}
\end{center}

\subsection{Scintillation Time Scale}

The scintillation time scale is the time for features in the diffraction 
pattern
to transport across the line of sight (LOS), combined with any reorganization
of the diffraction pattern itself.   These two contributions are determined
by the velocities of the source and observer and any bulk motion of
the intervening material, which change the sampling
geometry of the diffraction pattern, combined with
random velocities in the medium.
Traditionally the scintillation time scale is calculated as the 
$e^{-1}$ width along the time lag axis of the two dimensional
intensity correlation function,
$C(\delta\nu,\tau) = \langle I(t,\nu) I(t+\tau, \nu+\delta\nu)\rangle$.
For strong pulsars with steady pulse emission and scintillation
time scales of minutes or longer,  $C(\delta\nu,\tau)$ can be calculated
on a uniform grid of $\delta\nu$ and $\tau$.  However, for the Crab
pulsar,  the giant pulses allow only sporadic sampling along the
time-lag axis.  
At most of our observing frequencies, it is difficult to establish
values for $\dtd$ either because the DISS frequency structure is
unresolved (e.g. at 0.43 GHz) or because we cannot find enough close pairs
of giant pulses having adequate S/N to estimate reliably  the correlation
coefficient of the spectra.
However, at 1.475 GHz and 2.33 GHz, detectable pulses
allow us to estimate $\dtd$.

\smallskip
\begin{center}
\epsfxsize=\hsize
\epsfbox{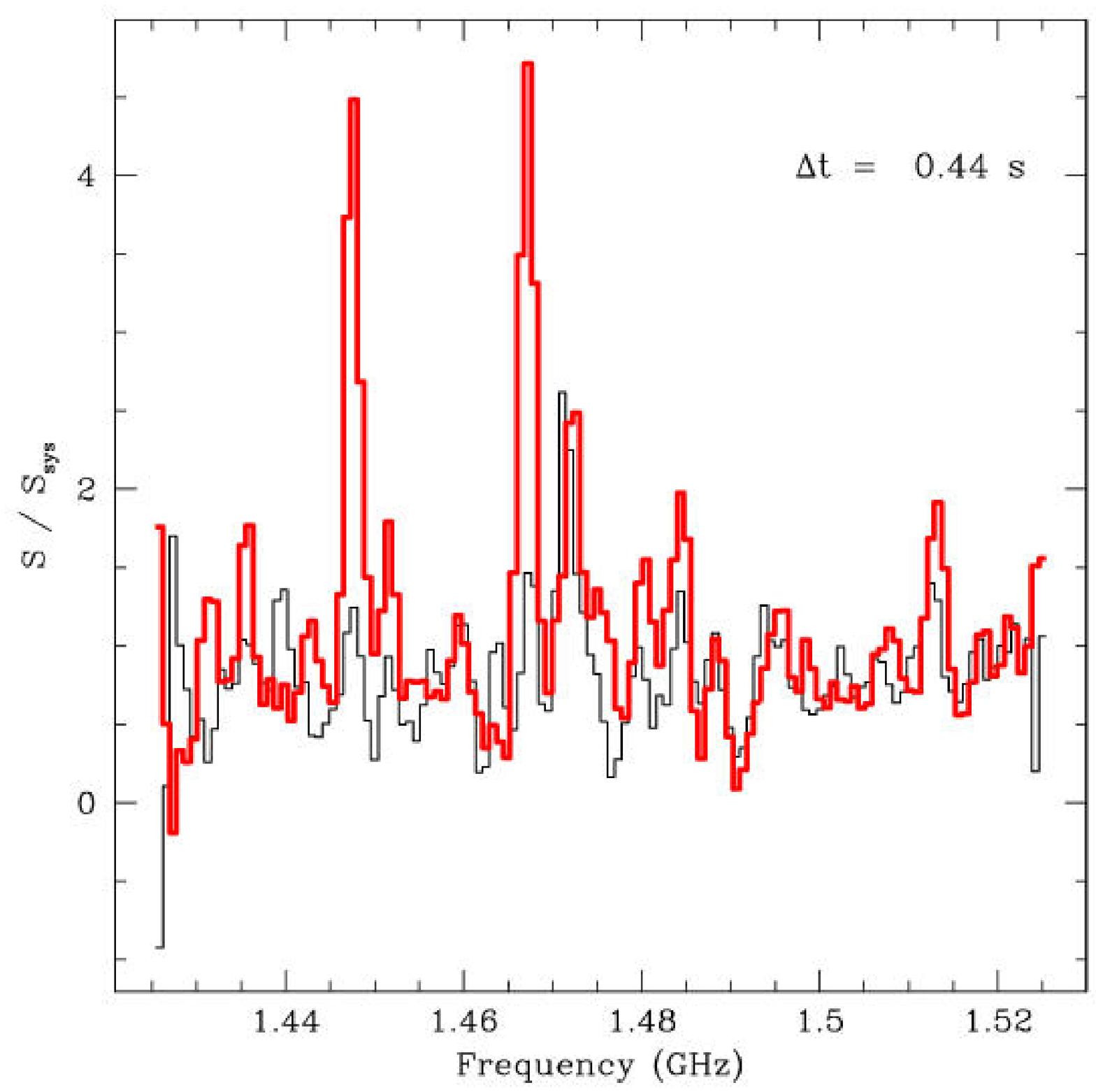}
\figcaption{
\label{fig:pair.spectra.1475MHz}
Plot showing spectra for two giant pulses spaced by
$\Delta t = 0.44$ sec and having S/N of 66 and 82 
(defined as peak to rms in the dedispersed time series)
for the curves with the light and heavy lines, 
respectively.
The largest spectral peaks in the stronger pulse
have S/N $\approx 100$.
Additive radiometer noise therefore contributes very little
to the spectral structure shown here. 
Features in the spectra for the two pulses  align
in frequency but have very different amplitudes.   The
correlation coefficient between the two spectra 
is 0.46.    As discussed in the Appendix, some of the frequency
structure is caused by DISS while other structure is associated
with the intrinsic noise properties of the pulse. 
}
\end{center}

Figure~\ref{fig:pair.spectra.1475MHz} shows spectra
for a close pair of high S/N pulses at 1.475 GHz.   While some of
the features in the spectra align, it is clear that the
frequency structure has decorrelated significantly.   The 
correlation coefficient is only 0.46.   If scintillations
were the only source of frequency structure, this would
imply an exceedingly short decorrelation time.  However
some of the frequency structure is associated with
the intrinsic noise of the pulsar signal.  
(Little structure is associated with additive radiometer noise
owing to the high S/N of 66 and 82 for the dedispersed pulses.)
The intrinsic
frequency structure has a frequency scale
$\Delta\nu_i \approx W_A^{-1}$, where $W_A$ is the
characteristic pulse width.   With
$W_A \approx 100\,\mu$s,  the intrinsic structure
is expected to show $\Delta\nu_i \approx 10$ kHz, much
narrower than the channel bandwidth of 0.78 MHz.
However,  substructure within the pulse envelope
comprising short-duration pulses of duration
$1 \, \mu s$ or less would increase this scale to 1 MHz or more.
Hankins \etal\ (2003) have identified substructure in giant
pulses on these short time scales.
We conclude that the intrinsic pulse structure is
responsible primarily for the fast  decorrelation 
between the pair of pulses. 
This conclusion is corroborated by a statistical study of a
large number of pulse pairs.

Figure~\ref{fig:ccfpairs.1475MHz} shows the correlation
coefficient $C(0, \tau)$ between a large number of pulse pairs
plotted against time separation, $\tau$.   
We have used only those pulses with S/N $> 20$ in the
dedispersed pulse in order to reduce scatter in the correlation
estimates from additive noise.   
The roll-off of the correlation coefficient
at larger lags represents a correlation time,
$\dtd \approx 25\pm 5$ sec at 1.475 MHz.   From Appendix~\ref{app:scints},
we expect the asymptotic correlation coefficient 
(at small lags) to be $1/(2+d_{\rm p}^2)$, 
in the mean, 
under the scintillating amplitude-modulated shot-noise (SAMPSN)
model  and where $d_{\rm p}$ is the degree of polarization ($\le 1)$.
This level is consistent with the level of correlation seen
at lags $\tau \lesssim 1$ sec if the pulses are typically 
highly polarized.
Consistency of giant pulse spectral statistics with 
SAMPSN implies that, typically, pulses at 1.475 GHz are composed of
a large number ($\gtrsim 5)$ of individual shot pulses in order
that the intrinsic fluctuations are Gaussian and thus contribute
to the rapid decorrelation seen (see Appendix).

A similar analysis for 2.33 GHz data yields a somewhat longer
time scale (Table~\ref{tab:scints}).  At still higher frequencies,
there are insufficient pairs of strong pulses to establish
the correlation time, though we are able to estimate the scintillation
bandwidth up to 3.5 GHz.   At 8.8 GHz, where the scintillation 
bandwidth is larger than the observation bandwidth,  we expect
spectral modulations to derive solely from amplitude-modulated
noise statistics (combined with radiometer noise), 
implying that the modulation index
$m_I = \sigma_I / I = \sqrt{(1+d_{\rm p}^2)/2}$ 
(after correction for any contribution
from radiometer noise, which is negligible for the largest
pulses).   For the pulse displayed in 
Figure~\ref{fig:tfplot.8.8GHz}, the modulation index is only
0.29.   A low modulation index suggests that the giant pulse of
that figure is dominated by a single shot pulse with duration
comparable to the reciprocal bandwidth, $\sim 10$ ns, or that
the giant pulse comprises a cluster of shot pulses with a similar
width.   Such results are not inconsistent with those of
Hankins \etal\ (2003), who found nanosecond structure within
individual giant pulses at 5 GHz.


\smallskip
\begin{center}
\epsfxsize=\hsize
\epsfbox{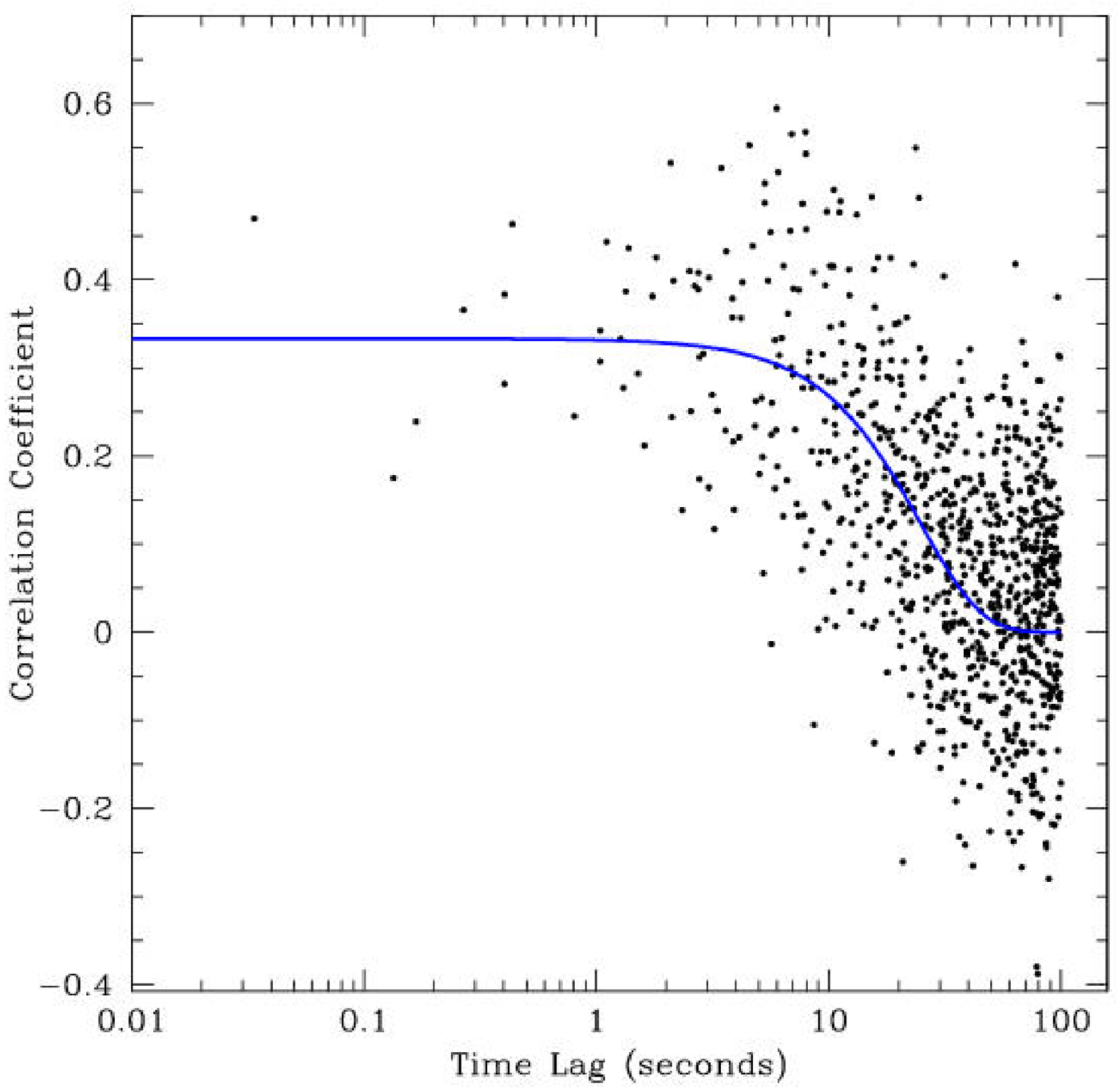}
\figcaption{
\label{fig:ccfpairs.1475MHz}
Plot of the correlation coefficient $C(\delta\nu=0,\tau)$ vs. time lag
$\tau$ between spectra of giant pulses at 1.475 GHz, where
$\tau$ is the time separation between pairs of giant pulses.  
At this frequency,
the number of close pairs is sufficient to establish that frequency
structure in the spectra is correlated over short time scales,
with an $e^{-1}$ time scale $\approx 25\pm 5$ s.  
The smallest lag 
occurs at one of the quantized values determined by the spacing of
the mainpulse and interpulse components.   
The line represents the mean correlation function expected
for scintillating amplitude modulated shot noise with a correlation
time of 25 sec.
}
\end{center}

\subsection{Scintillation Speed}

The effective speed with which the intensity pattern
caused by multipath propagation crosses the LOS
can be estimated from the scintillation bandwidth and time scale.
If we assume that electron density
fluctuations have a Kolmogorov spectrum and fill the LOS
uniformly, the pattern speed is (CR98)
\be
\vissKolu = \aissKolu
\frac{\sqrt{D \Delta \nu_{\rm d}}} {\nu \Delta t_{\rm d} },
\label{eq:vissKolu}
\ee
with $\aissKolu = 2.53 \times 10^4$ \kms\ for
$\nu$ in GHz,
$\Delta t_{\rm d}$ in s,
$\Delta \nu_{\rm d}$ in MHz
and $D$ in kpc.
We evaluate $\vissKolu$ at 1.475 GHz by using the decorrelation time
estimated in the previous section but by scaling the scintillation
bandwidth from 2.33 GHz, since it is unresolved at
1.475 GHz.   Scaling from 2.33 GHz using $\dnud\propto \nu^{4.4}$,
we obtain $\dnud \approx 0.31\pm 0.05$ MHz.  Using $D=2$ kpc for the
distance to the Crab Nebula, the scintillation speed is 
$\vissKolu \approx 540$ \kms.   Under the assumptions leading
to this estimate, $\vissKolu$ should be approximately equal to the 
transverse pulsar speed,
$\vert \vpperpvec \vert \approx 171\pm28$ \kms\
from an HST proper motion measurement
(Caraveo \& Mignani 1999); instead, there is a 
factor of 3 discrepancy.   

No element 
in the scattering geometry (pulsar, medium, or source) has a velocity
as large as $\vissKolu$ and it is reasonable to conclude that the
scattering medium is in fact not uniform along the LOS,
receiving contributions
from the Crab Nebula's filaments. 
First, we ignore the effects of the general ISM and consider filamentary
scattering screens at a distance $\ds \approx 1$ pc from the pulsar. 
Then, using Equations 13-18 of CR98 to correct for the 
geometric leveraging effects of screen(s) near the pulsar, we obtain 
\be
\viss   
&=& \wc\,\left[ \frac{2(D-\ds)}{\ds} \right ]^{1/2} \vissKolu  
	\label{eq:vissu2viss} \\
&\approx& 3.4\times 10^4\, 
\left( \frac{D/\ds}{2000}\right)
{\rm km \, s^{-1}}, \nonumber
\ee
where we have used $\wc \approx 1.05$ (c.f. Figure 1 of CR98).

For a thin screen, the pattern speed is related physically to the velocities of
the pulsar, observer, and medium  as (CR98, equation 4)
$\viss = (D/\ds) \vert \veffperpvec \vert$, or,
\be
\viss = 
 \left \vert 
 \left (\frac{D}{\ds} - 1\right)\vpperpvec + 
 \vobsperpvec -
 \left(\frac{D}{\ds}\right) \vismperpvec 
 \right \vert,
\label{eq:veffperpvec}
\ee
where $\vpperpvec, \vobsperpvec, \vismperpvec$ are the 
transverse velocities of
the pulsar, observer and medium relative to the local
standard of rest.  
The pulsar and medium's velocities are
``boosted'' by the factor $D/\ds \approx 2000$, so we may ignore
the Earth's motion in the following.

\smallskip
\begin{center}
\epsfxsize=\hsize
\begin{picture}(0,0)%
\includegraphics{fig13.pstex}%
\end{picture}%
\setlength{\unitlength}{2072sp}%
\begingroup\makeatletter\ifx\SetFigFont\undefined
\def\x#1#2#3#4#5#6#7\relax{\def\x{#1#2#3#4#5#6}}%
\expandafter\x\fmtname xxxxxx\relax \def\y{splain}%
\ifx\x\y   
\gdef\SetFigFont#1#2#3{%
  \ifnum #1<17\tiny\else \ifnum #1<20\small\else
  \ifnum #1<24\normalsize\else \ifnum #1<29\large\else
  \ifnum #1<34\Large\else \ifnum #1<41\LARGE\else
     \huge\fi\fi\fi\fi\fi\fi
  \csname #3\endcsname}%
\else
\gdef\SetFigFont#1#2#3{\begingroup
  \count@#1\relax \ifnum 25<\count@\count@25\fi
  \def\x{\endgroup\@setsize\SetFigFont{#2pt}}%
  \expandafter\x
    \csname \romannumeral\the\count@ pt\expandafter\endcsname
    \csname @\romannumeral\the\count@ pt\endcsname
  \csname #3\endcsname}%
\fi
\fi\endgroup
\begin{picture}(5344,6609)(1739,-6333)
\put(3691,-2401){\makebox(0,0)[lb]{\smash{\SetFigFont{8}{9.6}{rm}{\color[rgb]{0,0,0}${\mathbf{V_w}}(t+\tau)$}%
}}}
\put(4906,-3391){\makebox(0,0)[lb]{\smash{\SetFigFont{8}{9.6}{rm}{\color[rgb]{0,0,0}${\mathbf{{V_p}_{\perp}}}t$}%
}}}
\put(4816,-4291){\makebox(0,0)[lb]{\smash{\SetFigFont{8}{9.6}{rm}{\color[rgb]{0,0,0}${\mathbf{V_p}}t$}%
}}}
\end{picture}%

\figcaption{
Geometry showing the line of sight to the present-day location
of the pulsar, which has moved a distance 
${\mathbf{V_p}}t$ over the time $t$ since the supernova 
explosion, while the present day location of stellar
wind material along the line of sight is at a location
${\mathbf{V_w}}(t+\tau)$, taking into account an additional
time $\tau$ prior to the supernova explosion during which
the progenitor wind was active.   Under the assumption
of radial motion of the wind,  the wind material
(or blast-wave material condensed into filaments) has
a transverse speed
${\mathbf{V_w}}_{\perp} = \mathbf{V_p}_{\perp}/(1+\tau/t)$.
\label{fig:geometry}
}
\end{center}

The LOS filament velocities are 
$\lesssim 1500$ \kms \,(Fesen \& Kirshner 1982).
To estimate the effective transverse velocity, we need to consider
the symmetry of filament motions and how the geometry has changed
since the supernova explosion.  
Figure~\ref{fig:geometry} shows the geometry under the assumption
of purely radial motion by wind material (which may represent a
pre-supernova wind or blast-wave material).
The present-day transverse velocity
of the pulsar implies transverse filament velocities
$\vfilperpvec \equiv \vpperpvec$ (for filament segments along
the present-day LOS to the pulsar) 
if filaments originally had strictly
radial velocities (relative to the explosion center).  In this case,
$\Deltavpsperpvec \equiv \vpperpvec - \vfilperpvec$ 
vanishes and the effective transverse speed is merely the pulsar speed. 

However, nonradial filament motions are expected because the 
pulsar's progenitor star was rotating. Even with fairly slow
rotation, nonradial filament speeds of a few
\kms\ are boosted
by the factor $\ds/D$ to several thousand \kms.  Faster rotation
yields accordingly faster transverse filament speeds today.   Filaments
may also have arisen from material in or deflected by a circumstellar
disk around the progenitor (Fesen, Martin, \& Shull 1992), again
yielding large nonradial filament speeds today.   
Alternatively,  the material responsible for scattering in the Crab
Nebula could derive from the presupernova wind. 
In any case, we assume that $\Deltavpsperp$ is nonnegligible. 
It is not clear if nonradial filamentary motions have
been detected directly or not in the Crab Nebula (MacAlpine \etal\ 1994;
Schaller \& Fesen 2002).  However,  indirect constraints
(Trimble 1968; see also discussion in Backer, Wong \& Valanju 2000)
based on filament motions with respect to the explosion center
allow nonradial motions $\sim 70$ km s$^{-1}$.

By equating Eq.~\ref{eq:vissu2viss} and \ref{eq:veffperpvec},
we derive a constraint (independent of the distance to the Crab
Nebula),
\be
\label{eq:constraint}
\Deltavpsperp &=& \sqrt{2}\wc \vissKolu 
	\left( \frac{\ds}{D}\right)^{1/2} \\ 
	&\approx& 18 \ds^{1/2} \, {\rm km \, s^{-1}}. \nonumber
\ee
For pulsar-filament distances $\ds = 1$ pc, transverse filament speeds
relative to the pulsar
$\approx 18$ \kms\ are needed. 
We conclude that filaments that affect
the pulsar's radio emission possess modest nonradial motions
relative to the explosion center.

Alternatively, we can consider the combined effects of filaments
in the Crab Nebula and the general ISM.   Calculating the weighting
factor $\wdiss$ that relates $\viss$ to $\vissKolu$ 
(c.f. Eq. 10-18 of CR98) while taking into account a uniform ISM combined with
a thin screen, we find an expression analogous to that in
Eq.~\ref{eq:constraint},
\be
\label{eq:constraint2}
\Deltavpsperp &\approx& (3/8)^{3/5} \wc \vissKolu 
\left( \frac{\SM_{\rm ISM}}{\SM_{\rm CN}}\right)^{3/5}, 
\ee
where $\SM_{\rm ISM}$ and $\SM_{\rm CN}$ are the scattering measures
for the general ISM and the Crab Nebula, respectively, and we have
assumed that $\left(\ds/D)\right)\SM_{\rm ISM} \gg \SM_{\rm CN}$
but that $\SM_{\rm CN}>\SM_{\rm ISM}$.  The scattering measure
is the LOS integral of $\cnsq$, the spectral coefficient for electron density
irregularities (e.g. Cordes \& Lazio 2002).   Using values inferred
for the two scattering measures (from, e.g., the electron density
model, NE2001, of Cordes \& Lazio 2002 and from pulse broadening
of the Crab Nebula), Eq.~\ref{eq:constraint2} yields an estimate
$\Deltavpsperp\approx 20$ km s$^{-1}$ similar to that using
Eq.~\ref{eq:constraint}.


\section{Detectability of Giant Pulses from Extragalactic Crab-like Pulsars} \label{sec:detect}

The pulses we have identified are sufficiently strong to be detected
from other galaxies.    If we were to place the Crab pulsar in another galaxy,
the inverse square law would lessen pulse amplitudes but so too would it
decrease the contribution to the system temperature from the Crab Nebula. 
Consider a Crab-like pulsar in a Crab-Nebula-like nebula at
distance $\DNeb$.    The system noise level for this object is
(assuming it to be unresolved)
\be
\Ssys = \Ssyso + \left(\frac{\DCN}{\DNeb}\right)^2\SCN.
\ee 
The nebular contribution to the system noise becomes less than the
nominal system noise if $\Ssys < (1+\epsilon)\Ssyso$  or
\be
\DNeb > \DCN \left ( \frac{\SCN}{\epsilon \Ssyso}    \right )^{1/2}.
\label{eq:DNeb}
\ee

Table \ref{tab:dmax}  shows values of $\DNeb$ 
that satisfy inequality~\ref{eq:DNeb} for the Arecibo Telescope,
the Green Bank Telescope (GBT), the VLA (and the Extended
VLA), the Allen Telescope Array, 
the Low-Frequency Array (LOFAR), 
and the Square Kilometer Array (SKA).  
In all cases, we assume the nebula is unresolved.  For
the VLA, the ATA, LOFAR and some designs for the
SKA, this assumption will break down.
For $\epsilon = 1$ (equal contributions to $\Ssys$ from the nebula
and from receiver and background noise),  the Nebula is unimportant for objects
in the Magellanic Clouds for either of the existing telescopes.
However, for the SKA, nebular noise is dominant for such objects.
For the largest nearby spiral galaxies (M31 and M33), however,
nebular noise is negligible for all existing and contemplated telescopes.

The optimal frequency can also be determined. If we assume that the
spectrum is the same as that of the Crab pulsar,   lower frequencies
are favored unless propagation effects smear the pulse.   For the
Crab pulsar itself, 0.43 GHz is approximately the lowest frequency
at which propagation effects are sufficiently small.  For pulsars in
M31 or M33, the dispersion measures expected given their respective
Galactic latitudes and  inclinations are approximately equal to
the DM of the Crab pulsar.  Similarly, the scattering is expected
to be approximately the same.   Consequently,  we can use our 
0.43-GHz results on the Crab pulsar to estimate the $S/N$ expected
for extragalactic emitters of giant pulses. 

The strongest pulse observed at 0.43 GHz has 
$S/N_{\rm max} = 1.1\times 10^4$ even 
with the system noise dominated by the Crab Nebula.   For objects
in M31 ($D\approx 0.8$ Mpc) or further, the system noise is essentially
unaffected by the nebular contribution.    If the Crab pulsar were
not embedded in its nebula, the S/N of our largest pulse would have
been $S_{\rm CN} / \Ssyso \approx   300$ times larger, or
$3.3\times 10^6$.
For this particular pulse, the maximum distance it could be detected 
at a specified signal-to-noise ratio, $(S/N)_{\rm det}$ is
\be
D_{\rm max} 
	&=& \DCN 
	\left [
		\frac {(S/N)_{\rm max}} { (S/N)_{\rm det}}
		\left ( 1 + \frac{\SCN}{\Ssyso} \right)
		- \frac{\SCN}{\Ssyso}
	\right ]^{1/2} \\
	&\approx&  1.6\,{\rm Mpc}
	\left[\frac{(S/N)_{\rm det}}{5}\right] ^{-1/2}.
\ee
The largest 0.43 GHz pulse would thus be detectable from 
M33 ($D \approx 0.93$ Mpc) using  
the Arecibo telescope and our current spectrometer  at 
$S/N \approx 15$.   Using the GBT to observe M31 (since M31 is
outside the declination range of Arecibo), our
largest pulse would have $S/N \approx 4.8$.   Thus, a convincing
 detection of giant pulses from M31 with the GBT would require
longer dwell times than one hour in order that yet-stronger pulses
could be detected.   Detection of giant pulses is discussed in
general in Cordes \& Mclaughlin (2003) and in particular from nearby
galaxies in McLaughlin \& Cordes (2003).

LOFAR (the Low-frequency Array) would allow detection of a giant
pulse at 0.2 GHz at the 5$\sigma$ level out to a distance of 1.5 Mpc
for our largest pulse at 0.43 GHz, scaled to 0.2 GHz.
The SKA would yield $D_{\rm max} \approx 5.9$ Mpc for
a $5\sigma$ detection of  our largest pulse at 0.43 GHz. 
There are approximately 16 galaxies (of M33's size or larger)  
within this distance.  If pulsars like the Crab pulsar persist
in emitting giant pulses for a time of order the current
age of the Crab ($\sim 10^3$ yr) and if the birth rate of pulsars
scales as the ratio of a galaxy's mass to the Milky  Way's mass,
$\dot N_{\rm psr} \approx 10^{-2} \, {\rm yr^{-1}}\,
M_{\rm gal} / M_{\rm MW}$,  we expect that there should be a few
to about 20 such pulsars in each of these galaxies.   
Of course, giant-pulses are also emitted by millisecond pulsars
whose magnetic fields at their light cylinders are comparable to
that of the Crab pulsar (Cognard \etal\ 1996; Johnston \& Romani 2002)
so the numbers of detectable objects may be larger.   At present, however,
the Crab pulsar emits the most luminous giant pulses of any of these
objects and best serves as a prototype for detectable objects
from other galaxies. 

\section{Summary \& Conclusions}\label{sec:summary}

We have shown that giant pulses from the Crab pulsar are restricted
to only two of the pulse components seen in long-term average
profiles.
These components are the same as those detected at
optical, X-ray and gamma-ray energies, suggesting that the mechanism for
giant-pulse emission occurs high in the magnetosphere, where these
emissions are expected to originate. The occurence of giant pulses is
strongly frequency dependent.   We find that giant pulses `follow' 
the interpulse in pulse phase as it shifts to earlier phases above 
$\sim 4$ GHz.  We therefore conclude that the same physical region
produces both the low-frequency and the shifted, high-frequency
interpulse.   While the main pulse is dominant from 0.43 to
5.5 GHz, both in the average profiles and in the number of giant
pulses, at 8.8 GHz, the interpulse is dominant.   We have no
clear interpretation of this trend other than the usual suspect
processes: beaming and spectral dependence.   It is our aim to
analyze the profile shapes and giant-pulse occurrence histograms
along with multiwavelength pulse profiles extending to
$>100$ MeV gamma-rays in order to better constrain the roles
of beaming and coherence mechanisms.  This work will be deferred
to another paper.

Epoch dependence of the giant-pulse rate derives from
scintillation effects that appear to be strongly influenced by
plasma in the Crab Nebula. Backer, Wong \& Valanju (2000) demonstrate
that multiple images occur owing to the passage across the line of sight
of refracting plasma.  We establish the scintillation time scale
that is sufficiently short ($\sim 25$ s at 1.48 GHz) that plasma relatively
near the pulsar (i.e. inside the Crab Nebula) is required.
Our analysis on giant-pulse amplitudes suggests that, to the extent that
the Crab pulsar serves as a prototype of giant-pulse emission,
giant pulses from extragalactic pulsars should be detectable out
to large distances $\sim 1.6$ Mpc at Arecibo with existing 
back-end spectrometers.

We thank Bill Sisk and Jeff Hagen for developing the WAPP system
at the Arecibo Observatory, which was crucial for  providing
the data analyzed in this paper.  We thank Mal Ruderman for
helpful discussions.
NDRB is supported by an MIT-CfA Postdoctoral Fellowship at 
Haystack Observatory.
Work at Cornell University was 
supported by NSF grants AST 9819931 and 0206036.
MAM was also supported by an NSF MPS-DRF Fellowship. 
This work was also supported 
by the National Astronomy and
Ionosphere Center, which operates the Arecibo Observatory under
a cooperative agreement with the \hbox{NSF}.  THH thanks
NAIC for partial sabbatical leave support during this work.

\begin{deluxetable}{ccrcrcrr}
\tablecolumns{8}
\tablewidth{0pc}
\tablecaption{Observational Parameters \label{tab:params}}
\tablehead{
\\
\colhead{$\nu$} & 
\colhead{MJD} & 
\colhead{$T$\tablenotemark{a}} & 
\colhead{$B$} & 
\colhead{$\Delta t$} & 
\colhead{$\Delta \nu$} & 
\colhead{$\Delta t_{\rm DM}$}  & 
\colhead{$S_{\rm sys}$\tablenotemark{b}}\\
\colhead{(GHz)} & & \colhead{(hr)} & \colhead{(MHz)} &  
 	\colhead{($\mu s$)} & \colhead{(MHz)} & \colhead{($\mu s$)} 
	& \colhead{(Jy)}
}
\startdata
0.430 & 52304 & 1.0 & 12.5 & 128 & 0.024 & 145 & 1262 \\	
1.180 & 52277 & 1.2 & 100  & 100 & 0.781 & 224 &  309 \\ 
1.475 & 52277 & 1.2 & 100  & 100 & 0.781 & 115 &  291 \\	
2.150 & 52304-52306 & 1.3  & 100 &  32 & 1.562 &  74 &   79 \\	
2.330 & 52315 & 1.0  & 100 &  32 & 1.562 &  58 &   78 \\	
2.850 & 52306 & 1.0  & 100 &  32 & 1.562 &  32 &   74 \\	
3.500 & 52398-52412 & 3.5  & 100 &  64 & 1.562 &  64 &   41 \\	
4.150 & 52295-52337 & 2.9  & 100 &  32 & 3.125 &  21 &   20 \\	
5.500 & 52336-52411 & 2.3  & 100 &  32 & 3.125 &   9 &   20 \\	
8.600-8.800 & 52398-52414 & 3.1  & 100 &  16 & 3.125 &   2 &   22 \\	
%
\enddata
\tablenotetext{a}{$T$ is the total time of analyzed data, 
whether the pulsar was detected or not.}
\tablenotetext{b}{$S_{\rm sys}$ includes the contribution from
the Crab Nebula that takes into account flux dilution by the
telescope beam.}
\end{deluxetable}

\begin{deluxetable}{crlcc}
\tablecolumns{7}
\tablewidth{0pc}
\tablecaption{Giant Pulse Amplitude and Timing Statistics\label{tab:gpstats}}
\tablehead{
\\
\colhead{$\nu$} & 
\colhead{$T$\tablenotemark{a}} &
\colhead{$\overline\phi_{\rm ip} - \overline\phi_{\rm mp}$} &
\colhead{$N_{\rm ip} / N_{\rm mp}$} &
\colhead{$\dot N_{\rm GP}$\tablenotemark{b} } \\ 
\colhead{(GHz)} & 
\colhead{(hr)} &  
\colhead{(cycles)} & 
&
\colhead{($s^{-1}$)} 
}
\startdata
0.430 & 1.0  &$0.4032\pm10^{-4}$ & 0.56& 3.3   \\
1.180 & 0.47 &$0.402\pm0.001$    & 0.05& 0.51 \\
1.475 & 0.58 &$0.402\pm0.001$    & 0.05& 0.31 \\
2.150 & 0.15 &$0.403\pm0.002$    & 0.07& 0.25 \\
2.330 & 0.15 &$0.403\pm0.002$    & 0.08& 0.17 \\
2.850 & 0.26 &$0.404\pm0.003$    & 0.05& 0.11 \\
3.500 & 1.27 &$0.402\pm0.002$    & 0.04& 0.12 \\
4.150 & 1.49 &$0.394\pm0.002$    & 0.10& 0.31 \\
5.500\tablenotemark{c} & 0.30 & \hfil --- & --- &   0.02 \\
8.800 & 1.42 &$0.380\pm0.001$ & 27    & 0.44 \\
\enddata
\tablenotetext{a}{$T$ represents the total time included in the
average profiles of Figure~\ref{fig:composite}, which represents
only the high-S/N and RFI-free subset of the overall data.}
\tablenotetext{b}{For frequencies $\gtrsim$ 3 GHz, the
number of detected giant pulses varies significantly because
of diffractive interstellar scintillation.   Refractive
scintillation will also alter the detection rate at all frequencies.
}
\tablenotetext{c}{At 5 GHz, there are too few interpulses detected
to allow meaningful estimates of the MP to IP phase offset and
number ratio.}
\end{deluxetable}

\begin{deluxetable}{crcr}
\tablecolumns{4}
\tablewidth{0pc}
\tablecaption{Scintillation Parameters \label{tab:scints}}
\tablehead{
\\
\colhead{$\nu$} & 
\colhead{$\dnud$} & 
\colhead{$\dtd$} &
\colhead{$N_{\rm GP}$} \\ 
\colhead{(GHz)} & 
\colhead{(MHz)} &  
\colhead{($s$)} 
}
\startdata
0.43 & $<0.024$     & $\ldots$& 100 \\ 
1.48 & $<0.8$       & $25\pm 5$             & 180\\
2.33 & $2.3\pm 0.4$ & $35\pm 5$             & 170 \\
2.85 & $7\pm2$      & $\ldots$             &  60\\
3.50 & $15\pm10$    & $\ldots$             &  15 \\
\enddata
\end{deluxetable}

\begin{deluxetable}{rcrcr}
\tablecolumns{5}
\tablewidth{0pc}
\tablecaption{Maximum Distance for Importance of Nebular Noise \label{tab:dmax}}
\tablehead{
\\
\colhead{Telescope} & \colhead{$\nu$} & \colhead{G} & \colhead{T$_{\rm sys_0}$} & \colhead{$\epsilon^{-1/2}$D$_{\rm Neb}$} \\
 & \colhead{(GHz)} & \colhead{(K Jy$^{-1}$)} & \colhead{(K)} & \colhead{(kpc)} \\
		  }
\startdata
Arecibo	& 0.43 & 15 & 60 & 30 \\
Arecibo	& 1.4  & 11 & 40 & 30 \\
GBT     & 1.4  &  3 & 30 & 20 \\
VLA     & 0.33 &  2 & 165 & $<5$\tablenotemark{a} \\
VLA     & 1.4  &  2.8 & 35 & $<2.3$\tablenotemark{a} \\
ATA     & 1.4  & 2.5  & 50 & 13 \\
LOFAR   & 0.2  & 34\tablenotemark{b} & 476 &20 \\     
SKA     & 1.4  & 200\tablenotemark{c} & 50 & 120 
\enddata
\tablenotetext{a}{The VLA numbers are for the D configuration
which yields the largest contribution to the system temperature
from the Crab Nebula.}
\tablenotetext{b}{The gain for LOFAR is that for an inner
core of antennas that represents 75\% of the total collecting
\tablenotetext{c}{The full gain of the SKA is used.  In actuality,
only a fraction of this gain is likely to be available for 
time-domain studies of pulsars.}
area.}
\end{deluxetable}

\newpage
\appendix
\section{A. Frequency Structure from Scintillating Amplitude-modulated,
Polarized Shot Noise}\label{app:scints}

Frequency structure in the radio spectrum of a single pulse
is caused by both the statistical properties of the 
pulsed radiation at the time of emission
and by the interference effects of multipath propagation.
A model that suffices to describe many aspects of pulsar radiation
is the amplitude modulated noise model (Rickett 1975) augmented
to include polarized shot-noise statistics (Cordes 1976): the
amplitude-modulated, polarized shot-noise (AMPSN) model.
Frequency structure of single pulses was discussed by
Cordes \& Hankins (1979) in terms of the AMPSN model for
B0950+08.  Here we amplify on their treatment to show the interplay
of intrinsic and interference effects on the frequency structure. 
We thus develop the scintillating amplitude-modulated-shot-noise
model (SAMPSN).

Recent results (Hankins \etal\ 2003) imply that the Crab pulsar's
giant pulses are indeed comprised of individual shot nano-pulses,
in conformance with the AMPSN model.   
Let $\epsilon_e(t)$ be the complex, narrowband electric field emitted at the pulsar
and selected by the receiving system and mixed to baseband (see, e.g.,
Rickett 1975; Cordes 1976).   
We consider, for now, just a single polarization channel.
For an individual giant pulse,
we describe $\epsilon_e$ as an ensemble of $N_s$ 
shot pulses having individual amplitudes, $a_j$, but 
(for simplicity) identical shapes, $\Delta(t)$: 
\be
\epsilon_e(t) = \sum_{j=1}^{N_s} a_j \Delta(t - t_j).
\ee
The width of $\Delta(t)$ is the reciprocal of the receiver bandwidth
used to form $\epsilon(t)$.   The corresponding shot pulse at the
original radio frequency  has width 
$\sim \nu_{\rm RF}^{-1} \sim 0.1$ to 2 ns for our data.
Each shot pulse reaches the observer along $N_p$ paths owing
to multipath propagation between the pulsar and Earth.
Each path has an associated time delay, $\delta t_p$, and
amplitude $g_k$.  The set of paths changes on a time scale
that we assume is much longer than a single spin period.
Also, all propagation quantities ($N_p, \delta t_p$, and $g_k$)
are strong functions of frequency because the refractive
index is that for a cold plasma.
Using the total propagation time,   
$D/c + \delta t_p$, the received electric field is
\be
\epsilon(t) \propto \sum_{j=1}^{N_s} a_j 
		\sum_{k=1}^{N_p} g_k \Delta(t - t_j - D/c - \delta {t_p}_k). 
\ee
Here we have ignored delays from dispersive propagation through the ISM
because they are deterministic and correctable.
Denoting a Fourier transform with a tilde and using the shift theorem, 
we find that the instantaneous spectrum is
\be
I(\nu) \equiv 
	\vert \tilde \epsilon(\nu) \vert^2 \propto 
	\vert \tilde \Delta(\nu) \vert^2
	A(\nu) 
	G(\nu), 
\ee
where $\nu$ is the baseband frequency and
\be
A(\nu) &=& \left \vert \sum_{j=1}^{N_s} a_j 
	e^{-2\pi i \nu t_j} \right\vert^2 \\
G(\nu) &=& \left \vert \sum_{k=1}^{N_p} g_j 
	e^{-2\pi i \nu \delta {t_p}_k} \right\vert^2.
\ee

In the limit of large $N_s$ and $N_p$, we expect Gaussian statistics
for the sums in the equations for
$A$ and $G$.   Consequently, $A$ and $G$ will both have exponential
statistics,  for which
$\langle A^2 \rangle / \langle A \rangle ^2 =
\langle G^2 \rangle / \langle G \rangle ^2 = 2$.
As is well known
in the scintillation literature (e.g. Rickett 1990),  scintillation
fluctuations in the strong-scattering regime have exponential
statistics for a point source if there is no bandwidth smoothing.
Note that our formulation of amplitude-modulated noise
differs from that of Rickett (1975) and Cordes (1976), who
model the emitted signal as $\epsilon_e(t) = a(t) m(t)$,
where $a(t)$ is an envelope function that modulates the noise
process $m(t)$.   Instead,  the envelope function is
absorbed into the particular distribution of emission times,
$t_j$.

To isolate the frequency structure of $A(\nu)$ from $G(\nu)$,
one must take into account their characteristic time scales.
It is reasonable to assume that the pattern of shot pulses in 
$\epsilon(t)$ does not repeat.  
On a physical basis, such shot pulses may result
from the sweeping of relativistic beams through the LOS
or they may represent {\it bona fide} temporal modulations.  Either
way, on time scales $\gtrsim P/2\pi$ ($P$ is the pulsar period),
we expect the relativistic plasma flow in the pulsar magnetosphere
to have reorganized completely.    The scintillation pattern,
on the other hand,  is sustained.   
It is usually true for pulsars that if the scintillation frequency structure
is resolved by the spectrometer, 
it persists over time scales of seconds to
hours, depending on the pulsar and frequency.  For heavily scattered
pulsars, the frequency structure is too fine to resolve and the
scintillation time is accordingly short. Thus, for most pulsars,
the frequency structure in $G(\nu)$, if resolved,  is characterized
by averaging $I(\nu)$ over many individual pulses  and then performing a 
correlation analysis to determine the characteristic bandwidth.

For the Crab pulsar, which emits giant pulses only sporadically,
it is more difficult to separate $A(\nu)$ from $G(\nu)$  and also
estimate the scintillation time scale.   For an individual
giant pulse,  $A(\nu)$ and $G(\nu)$ both contribute to the
observed frequency structure and with similar statistics.   However,
the characteristic width of $G(\nu)$ scales strongly with frequency,
as discussed above, while $A(\nu)$ is associated with the temporal
widths of the giant pulses  and may be less frequency dependent. 

\subsection{Statistics for a Single Polarization Channel}

Some useful statistics of the SAMPSN model are as follows.
The modulation index of the spectrum $I(\nu)$
is $\sigma_I / \langle I\rangle = 3$ when $A$ and $G$ both have
exponential statistics.  For a pair of pulses for which
$A(\nu)$ has decorrelated completely while $G(\nu)$ is perfectly
correlated, we expect the cross correlation to
be 
$\rho_{12} = 
\langle \delta I_1(\nu) \delta I_2(\nu) \rangle /
\sigma_I^2 = 1/3$.
The correlation coefficient will decline to zero on  a time lag between
the pair of pulses determined by the characteristic scintillation time, 
defined as the lag at which the correlation coefficient is $e^{-1}$ of
its maximum value of 1/3.

The autocorrelation function (ACF) of the spectra for single pulses can
be written in the form
\be
R(\delta\nu) = \langle I(\nu) I(\nu+\delta\nu)\rangle
	= R_{\vert \tilde \Delta \vert^2}(\delta\nu)
	  R_A(\delta\nu)
	  R_G(\delta\nu).
\ee
$R_{\vert \tilde \Delta \vert^2}(\delta\nu)$ is a broad function that
is the ACF of the bandpass filter used to form $\epsilon(t)$, while
$R_A$ and $R_G$ can be much narrower and are of the form
$R_X(\delta\nu) = \langle X \rangle^2 [1 + m_X^2\rho_X(\delta\nu)]$, 
where $m_X = 1$ for exponential
statistics and $\rho_X(0) = 1$.  
The intensity correlation function,
\be
R(\delta\nu) = 
	\langle G \rangle^2
	\langle A \rangle^2
	R_{\vert \tilde \Delta \vert^2}(\delta\nu)
	[1 + m_A^2\rho_A(\delta\nu)]
	[1 + m_G^2\rho_G(\delta\nu)],
\ee
will typically have a narrower component and a broader component associated
with $\rho_A$ and $\rho_G$, respectively, or vice versa.  
The total squared-modulation index  is
$m^2 = \sigma_I^2 / \langle I \rangle^2 = 
R(0)/\langle G \rangle^2 \langle A \rangle^2 R_{\vert \tilde \Delta \vert^2}(0) 
-1 = 1+m_A^2+m_G^2 = 3$.   
If the data are channelized with channel bandwidths
larger than the characteristic bandwidth of $A$ of $G$, the 
modulation index will be reduced.

Some giant pulses comprise a small number of shot pulses,
$N_s \approx$ a few, in which case $A(\nu)$ will have non-exponential
statistics.  For example, for two equal-amplitude
shot pulses separated by $\Delta t_{12}$,
$A(\nu)\propto (1 - \cos2\pi\nu\Delta t_{12})$.   If
$\nu\Delta t_{12} \gg 1$, we would have $m_A = 2^{-1/2}$ and the
spectral shape would be oscillatory.    
In the limit of a single shot pulse (or a cluster of shot pulses
contained within an interval smaller than the reciprocal bandwidth),
the modulation across the bandwidth would derive solely from
$G$, the scintillation factor.

\subsection{Statistics for the Total Intensity}

When two polarization channels are summed to yield the total intensity,
as in the analysis of this paper, the statistics are altered.  
Scintillations are identical for the two polarizations while
the frequency structure from the AMPSN will differ according to
the degree of polarization.  If the signal is 100\% polarized,
the total intensity will have the same statistics as that of a single
polarization channel containing the signal, while for an unpolarized
signal, the AMPSN spectral fluctuations will be reduced by $\sqrt{2}$.

Letting $d_{\rm p}$ equal the total degree of polarization (linear and circular),
it may be shown (Cordes 1976) that the intensity modulation index is
now (in the limit of Gaussian statistics for a single polarization
channel)
\be
m_I^2  = 
m_G^2 + (1 + m_G^2)(1+d_{\rm p}^2)/2 
\stackrel{m_G=1}{=} 2 + d_{\rm p}^2, 
\ee
where the last equality holds for $m_G=1$.   Now, when the total
intensity spectra of pulse pairs are cross-correlated, we have
\be
\rho_{12} = \langle \delta I_1(\nu) \delta I_2(\nu) \rangle / \sigma_I^2 
= \frac{m_G^2}{m_G^2 + (1 + m_G^2)(1+d_{\rm p}^2)/2}   
\stackrel{m_G=1}{=} \frac{1}{2+d_{\rm p}^2}.
\ee

\end{document}